\PassOptionsToPackage{square,comma,numbers,sort&compress,super}{natbib}
\documentclass[12pt]{article}
\usepackage{amsthm, amscd, amsfonts, amssymb, graphicx,tikz, color, environ}
\usepackage{amsmath}
\usepackage{hyperref}
\usepackage{caption} 
\usepackage{bm}
\usepackage{comment}
\usepackage{float}
\usepackage{cleveref}
\usepackage{ragged2e}
\usepackage[utf8]{inputenc}
\newtheorem{theorem}{Theorem}[section]

\newtheorem{lemma}[theorem]{Lemma}

\usepackage{enumitem}
\newcommand\numberthis{\addtocounter{equation}{1}\tag{\theequation}}
\usepackage[sort&compress,numbers,comma,square]{natbib} %for references
\hypersetup{
   colorlinks, % make the links colored
    linkcolor=blue, % color TOC links in blue
  citecolor=blue
}
\begin{document}
%\begin{document}
\begin{center}{\Large Likelihood-based Inference for Skewed Responses in a Crossover Trial Setup}\end{center}
%\vspace{.1em}
\begin{center}{{ Savita Pareek\textsuperscript{\color{blue}1}, Kalyan Das\textsuperscript{\color{blue}1}, and Siuli Mukhopadhyay\textsuperscript{\color{blue}1\color{black},} {\footnote[2]{Corresponding author. Email: siuli@math.iitb.ac.in}}}\\
%{\it Department of Mathematics, Indian Institute of Technology Bombay,}\\ {\it Mumbai 400 076, India}}
\vspace{.15em}
{\textsuperscript{1}\it Department of Mathematics, Indian Institute of Technology Bombay,}\\ {\it Mumbai 400 076, India}}
\end{center}
\date{}
\hrule
\vspace{.2em}
\section*{Abstract}
This work proposes a statistical model for crossover trials with multiple skewed responses measured in each period. A 3 $\times$ 3 crossover trial data where different doses of a drug were administered to subjects with a history of seasonal asthma rhinitis to grass pollen is used for motivation. In each period, gene expression values for ten genes were measured from each subject. It considers a linear mixed effect model with skew normally distributed random effect or random error term to model the asymmetric responses in the crossover trials. The paper examines cases (i) when a random effect follows a skew-normal distribution, as well as (ii) when a random error follows a skew-normal distribution. The EM algorithm is used in both cases to compute maximum likelihood estimates of parameters. Simulations and crossover data from the gene expression study illustrate the proposed approach.

\smallskip
\noindent \textbf{Keywords.} Crossover design, Mixed effect models, Skew-normal distribution, EM algorithm.
\section{Introduction} 
Crossover design is a specific type of longitudinal study in which every subject receives different treatments in different periods. It is most commonly used in the pharmaceutical
industry and other medical fields to investigate the safety and efficacy of new drugs or treatments. In a crossover design, the treatment effect is compared within the subject since each subject acts as its own control. Therefore, fewer subjects are needed than in a parallel design in order to achieve the same level of accuracy. An in-depth description of crossover trials can be found in the books by \citet{senn2002}, \citet{Jones}. %The books by \citet{senn2002}, \citet{Jones} provide a detailed review of crossover trials. %For a detailed review of crossover trials, refer to the books by (\citet{senn2002};  \citet{Jones}).  

In many clinical studies, we sometimes encounter crossover trials with measurements on two or more response variates. For example, one may consider the measurement of both systolic (SBP) and diastolic (DBP) blood pressure of subjects in each period (\citet{Grender1993}) or blood sugar levels recorded at multiple time points in each period (\citet{Putt1999}) or microarray gene expression profiles of subjects measured in each period (\citet{Leaker2016}). Other than multiple responses in some real-life crossover trials, such as bioequivalence trials, we come across multi-modal or skewed responses (\citet{Jones}). One solution to such cases is to apply the log or box-cox power transformation. However, this may not be a reasonable solution for most cases, thus causing a considerable modeling challenge for the statistician (\citet{logtransform}). The use of Bayesian methods based on extended generalized gamma distribution and skew-t distribution has been discussed in the literature for such bioequivalence studies with the skewed response (\citet{2016}, \citet{2021}). 
Using conventional analysis techniques for such skewed multivariate data may lead to an incorrect and biased parameter and variance estimates. 

In comparison to univariate responses, crossover trials with multiple responses measured in each period have  been addressed by very few researchers, namely (\citet{Grender1993}, \citet{chin1996}, \citet{Putt1999}, \citet{Tudor2000}, \citet{multivariate2007}, \citet{Pareek2021}). The main reason behind the scarcity of literature may be the difficulty in modeling the complex inter and intra-response relationships. The linear mixed or random-effects model (\citet{laird1982}) is a widely used technique for analyzing data from crossover studies. These models routinely assume both random effects and random errors to be normally distributed. While the assumption of normality is easy to execute, several authors (\citet{zhang2001}, \citet{ghidy2004}, \citet{drik2017} and \citet{drik2019}) have questioned the robustness of fitted models, specifically when data show multi-modality and skewness. \citet{zhang2001} have demonstrated that in the Framingham heart study, estimated subject-specific intercepts are not normally distributed, and the use of normal distribution in this scenario has resulted in less efficient inferences on intercept and treatment effect or subject level covariate. \citet{drik2017} has argued that maximum likelihood estimates of regression parameters may be biased when the distribution of random effects is not correctly specified. Therefore, it is of practical interest to develop statistical models having random effects or random errors to be skew-normal while allowing for multiple response measurements in each period. In the last decade, a substantial amount of work has been done on skewed responses in longitudinal studies. \citet{lachos2005} have developed an EM-type algorithm for maximum likelihood estimation in linear mixed effect (LME) models, assuming random effects or random errors to be multivariate skew-normal. \citet{lachos2010} proposed the longitudinal data modeling using linear mixed models with skew-normal independent (SNI) distribution for random effects and normal independent (NI) distributions of within-subject errors. SNI distributions are a sub-class of scale mixture of skew-normal (SMSN) distribution introduced by \citet{branco2001}. Recently \citet{pieri2019} have developed non-linear regression models assuming random effects to be a class of SMSN such as skew-normal, skew-t, skew-slash, and skew-contaminated normal distributions. \citet{schu2020} presented the maximum likelihood estimation using an EM-type algorithm for correlated error terms such as the auto-regressive correlation of order p in the LME model with SNI random effects. However, none of these authors have considered crossover trials with multiple and skewed measurements in each period.% skewed data in crossover trials has very limited literature.

In this work, we address the issue of skewed responses in a multivariate crossover setup. A mixed-effect model approach with an EM-based estimation method is proposed. We consider the linear mixed-effects model with skew-normally distributed random subject-specific effect or random error term. The hierarchical representation of the model makes it feasible to use the EM-type algorithm, which produces the closed-form expressions for E and M-steps for special cases. We further analyze the gene expression data (\citet{Leaker2016}) from a three-period three-treatment crossover design. 

The rest of the article is structured as follows: after a brief discussion on the gene expression case study in \Cref{case study}, the proposed random effects model is presented in \Cref{model}. In \Cref{mle sec}, a likelihood-based EM-type algorithm is considered for parameter estimation. In \Cref{sim study}, two simulation studies are conducted to examine the performance of parameter estimates. The proposed methodology's benefits are illustrated by analyzing a gene expression dataset in \Cref{gene sec}, while \Cref{comp} provides the relevant computational specifics. Finally, concluding remarks are provided in \Cref{conc}.

\section{Case Study: Multivariate Crossover Trial of Oral Prednisone}\label{case study}
We use  a gene expression dataset from (\citet{Leaker2016}) as a case study. The dataset is publicly available from the NCBI Gene Expression Omnibus (\citet{Clough2016}) and can be accessed using the hyperlink, \href{https://www.ncbi.nlm.nih.gov/geo/query/acc.cgi?acc=GSE67200}{nasal mRNA data}. In the gene expression study, results from a  randomized double-blind, placebo-controlled, three-period, crossover trial are considered to evaluate the effects of two single doses of oral prednisone (10 mg, 25 mg) on inflammatory mediators measured in nasal exudates after nasal allergen challenge in susceptible individuals with allergic rhinitis.
All subjects have a history of seasonal asthma rhinitis to grass pollen and a positive result from the intraepidermal skin prick test to grass pollen extract. Seventeen subjects were enrolled in the study and assigned to three treatment sequences/groups, out of which five subject observations with missing values were not considered.
The main interest here is to study the effect of treatments and genes on allergic reactions to grass pollen. The outcomes measured are a fold change of mRNA expression levels (pg ml$^{-1}$), \textit{i.e.,} changes in gene expression values for ten genes recorded in the nasal allergen challenge. Subjects with missing observations are excluded from our analysis. The study design is described in \Cref{tab:study design}. %Our analysis ignores the subjects with missing observations.
 
 \begin{table}[!htp]
\centering
\caption{Study design: a three-way crossover trial to examine the effects of a single oral
dose of prednisone (10 or 25 mg) versus placebo given before nasal allergen challenge (NAC).}\label{tab:study design}
\begin{tabular}{llll}
\hline
 & Period 1           & Period 2           & Period 3           \\
 \hline
Sequence 1 (4 subjects) & 10mg    & Placebo            & 25   mg  \\
Sequence 2 (4 subjects)& 25   mg  & 10   mg  & Placebo            \\
Sequence 3 (4 subjects) & Placebo            & 25   mg  & 10   mg 
\\ \hline
\end{tabular}
\end{table}

As an example, sequence 1 indicates that 10 mg prednisone is given to the subjects (subject 1 to subject 4) in period 1, followed by a washout, a placebo is given in period 2, and again after a washout, 25 mg prednisone is given in period 3. 

 Before model fitting, we ran some exploratory analysis on the gene data as follows: 
 \begin{itemize}
     \item[(i)] The density and normal Q-Q plot of the original and log-transformed responses as given in \Cref{fig1}. We determine if the given responses are representative of a normal population by using the Shapiro-Wilk tests (\citet{shapiro1965}). \Cref{tab2:norm} interprets raw and transformed responses' density, normal Q-Q plots, and Shapiro-Wilk tests at 5\% significance level. The observations made in \Cref{tab2:norm} indicate asymmetric behavior of the gene expression levels. Moreover, the Shapiro-Wilk test on box-cox transformed responses yields a p-value $<0.0001$, indicating that the box-cox power transformation also fails to conform the responses to normality.  %Log transformation is one of the most common and widely used transformations to tackle skewed data. However, in our data set, the histogram, normal QQ-plot (\Cref{fig1}, Plots C, D), and the Shapiro-Wilk test for normality (p-value $<0.001$) of log-transformed responses as well does not show typical normal behavior. 
     \begin{table}[!htp]
     \caption{Testing for normality using the density, Q-Q plot, and Shapiro-Wilk test for original and log transformed responses.}\label{tab2:norm}
\begin{tabular}{lll}
\hline
Method   & \multicolumn{1}{l}{Original responses}    & Log-transformed responses\\ \hline
\begin{tabular}[c]{@{}l@{}}Shapiro-Wilk test \\ (p-value)\end{tabular} & \multicolumn{1}{l}{\textless{}0.0001}     & \textless{}0.0001   \\ 
Density plot                                                              & \multicolumn{1}{l}{\begin{tabular}[c]{@{}l@{}}Does not resemble \\ a bell-shaped curve.\end{tabular}}                                                     & \begin{tabular}[c]{@{}l@{}}Does not resemble \\ a bell-shaped curve.\end{tabular}                                                     \\ 
Normal Q-Q plot                                                        & \multicolumn{1}{l}{\begin{tabular}[c]{@{}l@{}}Both ends of the Q-Q plot\\  deviates significantly \\ from the diagonal line.\end{tabular}} & \begin{tabular}[c]{@{}l@{}}Both ends of the Q-Q plot\\  deviates significantly \\ from the diagonal line.\end{tabular} \\ \hline
\end{tabular}
\end{table}

     \begin{figure}[ht]
\centering
  \includegraphics[width=12cm]{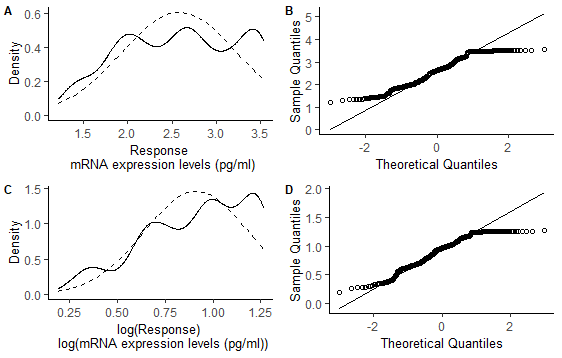}
  \caption{{Density and normal Q-Q plot comparisons: observed raw responses (Plots A and B) and log-transformed responses (Plots C and D), with true normal density indicated by a dashed line.}}
  %Density and normal Q-Q plots: Plots (A, B) corresponds to the observed raw responses. Plots (C, D) correspond to the log-transformed responses. Density plot with dashed lines shows true normal density.
  \label{fig1}
 \end{figure}
     \item [(ii)] The objective is to evaluate the effect of different genes on the relationship between period versus responses, treatment versus responses, and subject versus responses, respectively. This can be achieved by creating interaction plots. %that include period versus gene, treatment versus gene, and subject versus gene. %To see how the relationship between period or treatment or subjects and responses depends on the various genes, draw the period by gene, treatment by gene, and subject by gene interaction plots. 
    % \Cref{fig3} shows that plots A and B mostly have parallel lines. In plot C, there is a slight crossing between the lines of genes 2, 9, and 10. Therefore, periods by gene, treatments by gene, and subjects by gene interactions can be skipped.
    According to \Cref{fig3}, plots A and B primarily have parallel lines. Plot C shows a slight overlap between the lines for genes 2, 9, and 10. Nevertheless, our model does not consider subject versus gene interactions in order to reduce model complexity and parameters, but it may be possible to examine such interactions in the future.
      \begin{figure}[!htp]
\centering
  \includegraphics[width=1\textwidth,height=17cm,keepaspectratio]{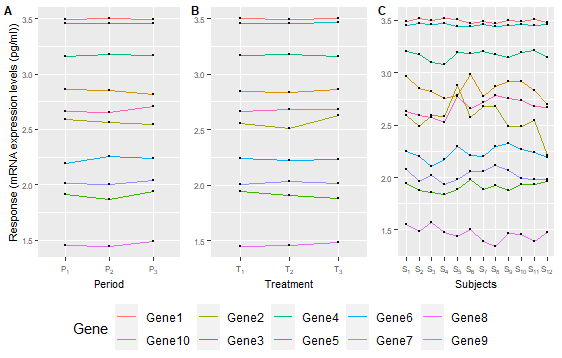}
  \caption{{ Interaction Plots: gene interactions by time period, treatment, and subject. Plot A represents gene interactions based on time periods (P$_i$, i=1, 2, 3), Plot B represents gene interactions based on treatments (T$_i$, i=1, 2, 3), and Plot C represents gene interactions based on subjects (S$_i$, i=1, 2, \ldots, 12). Each point in Plots A, B, and C displays the average response of a given gene (averaged over sequences).}}
  \label{fig3}
 \end{figure}
     \item [(iii)]
    As a further step, a normal linear mixed-effects model was fitted, with period, treatment, and gene as fixed effects and subject-specific normal random effects. As a result of fitting the model, we constructed a Q-Q plot of the estimated random intercepts, as well as a residual versus estimated responses plot. From \Cref{fig4}, we see that the estimated subject-specific intercept histogram, Q-Q plot, and the Shapiro-Wilk test (having p-value $0.33$) depict no apparent non-normal patterns. However, the normal Q-Q plot and the Shapiro-Wilk test (p-value $< 0.0001$) of the standardized residuals show asymmetric behavior, and the residual versus fitted values plot also indicates non-constant variance. %distributed.%of residuals standardized residuals vs. fitted values  To examine the existence of skewness in random error, we checked the residual plot of the model fitted in (ii). The residual vs. fitted values plot (\Cref{fig4}) and the Shapiro-Wilk test for normality (p-value $< 0.001$) indicate asymmetric behavior of residuals.
         \begin{figure}[!htp]
\centering
  \includegraphics[width=10cm]{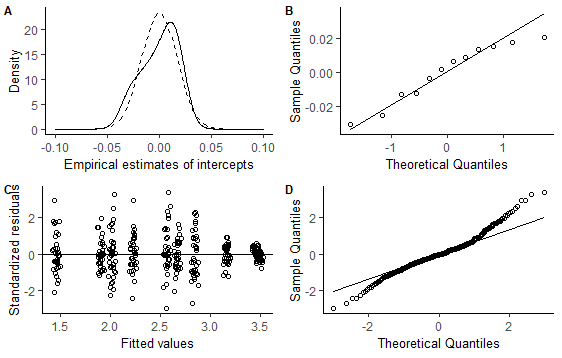}
  \caption{{Validation graphs for normal model fitting based on estimated random effects and standardized residuals: Subplots A and B represent the density and normal Q-Q Plot of the estimated random intercepts, while Subplots C and D depict the standardized residuals \textit{vs.} fitted values and normal Q-Q Plots of the standardized residuals, with the true normal density displayed as a dashed line.}}
  \label{fig4}
 \end{figure}
 \end{itemize}

 Based on the exploratory analysis, we propose the following skewed-normal model for the multiple gene responses.
% The most commonly used log transformation, box-cox power transformation, does not seem to be effective in tackling non-normal responses in the gene data set. So, we examine the original responses in the rest of the article. Also, considering that box plots are similar for various gene expression levels in \Cref{fig2}, it may be assumed that the period and treatment effects are the same for all the ten gene expression levels. It is also justifiable to omit the interaction plots for the period versus gene, treatment versus gene, and subject versus gene, as the lines in these plots tend to be mostly parallel (\Cref{fig3}). Therefore, in the following section, we propose a model that does not vary the treatment and period effects with multiple responses and excludes the various interaction terms. % In sum, we learn that for non-normal responses are non-normal, and fitting the above data with the proposed model of the next section having non-normal error distribution may provide a better solution. 
 
\section{Model and Notations}\label{model}
In this section, we propose a statistical model motivated by the gene expression data to fit multiple skewed responses measured in a crossover trial.
Suppose, $y_{i j t k}$ denotes the $k^{\text{th}}$ gene expression at $t^{\text{th}}$ time period for $j^{\text{th}}$ subject in $i^{\text{th}}$ sequence, where $i = 1,2, \ldots, s$;  $j = 1,2, \ldots, n_i$; $t = 1,2,\ldots, p$ and $k = 1,2, \ldots, m$. There are in total $n$ subjects, where $n=\sum_{i=1}^{s}n_i$.  We may write
\begin{equation}\label{mod eq}
y_{i j t k}=\mu+\pi_{t}+\tau_{d[i, t]}+g_{k}+s_{i j}+e_{i j t k},
\end{equation}
where, $\mu$ is the overall mean, $\pi_{t}$ is the $t^{\text{th}}$ period effect, $\tau_{d[i, t]}$ is the effect of treatment applied in $t^{\text{th}}$ period of $i^{\text{th}}$ sequence, $g_{k}$ is the $k^{\text{th}}$ gene effect, $s_{i j}$ is the subject-specific   random effect, and $e_{i j t k}$ is the random error.

For ease of exposition, we rewrite \cref{mod eq} in matrix form as
\begin{equation}\label{mat form}
\bm{y}_{ij}=\bm{ X}_{ij}\bm{\beta}+\bm{Z}_{ij}b_{ij}+\bm{e}_{ij};~ i=1(1)s,~ j=1(1)n_i.
\end{equation}
For fixed values of i and j,
\begin{itemize}
    \item[(i)] $\bm{y}_{ij}=(y_{ij11},\ldots, y_{ij1m},y_{ij21},\ldots, y_{ij2m},\cdots, y_{ijp1},\ldots, y_{ijpm})^{T}$ is a response vector of length $pm$.
  \item[(ii)] $\bm{X}_{ij}$ is the design matrix of order $pm\times (p+t+m-2)$ corresponding to the fixed effects, where, $$\bm{X}_{ij}=(\bm{1}_{pm},(\bm{0}_{m \times (p-1)},(\bm{I}_{p-1}\otimes \bm{1}_{m})^{T})^{T}_{pm \times (p-1)},\bm{T}_{pm \times (t-1)},\bm{1}_{p}\otimes(\bm{0}_{m-1},\bm{I}_{m-1})^{T}).$$
 The ${pm \times (t-1)}$ matrix of treatment effects is represented as $\bm{T}=(\bm{T}_{1}^{T}, \ldots \bm{T}_{p}^{T})^{T}$, where,  $ \bm{T}_{u}=(\bm{1}_{m} \otimes (a_{2},\ldots, a_{t}))_{m \times (t-1)}$ for $u= 1(1) p$, and $a_l$ for $l=2(1)p$ is an indicator variable which takes value $1$ if the $l^{\text{th}}$ treatment is assigned to the $u^{\text{th}}$ period and $0$ otherwise. 
\item[(iii)] $\bm{\beta}$ is the parameter vector of length $(p+t+m-2)$ corresponding to the fixed effects, $\bm{\beta}=(\mu , \pi_{2} ,\dots \pi_{p},  \tau_{2},\ldots,\tau_{t}, g_{2},\ldots,g_{m}).$
 %The parameters of interest are,  $\bm{\theta}=(\mu , \pi_{2} ,\dots \pi_{p},  \tau_{2},\ldots,\tau_{t}, g_{2},\ldots,g_{m}, \sigma_e^2,\sigma_s^2,\lambda_e,\lambda_s).$
 \item[(iv)] $\bm{Z}_{ij}=\bm{1}_{pm}$ is the design matrix corresponding to the random effects. Consider, for example, two responses being measured for each subject in every period of a $2 \times 2$ crossover design with treatment sequences AB and BA having five subjects in each. Then for $i = 1, 2; j= 1, 2, \ldots, 5$, the design matrices $\bm{X}_{i j}, \bm{Z}_{i j}$ and parameter vector $\bm{\beta}$ will be
$$\bm{X}_{1j}=  \begin{bmatrix}
1 & 0 & 0&0\\
1 & 0 & 0&1\\
1&1&1&0\\
1&1&1&1
\end{bmatrix},
\: \quad
\bm{X}_{2j}=  \begin{bmatrix}
1 & 0 & 1&0\\
1 & 0 & 1&1\\
1&1&0&0\\
1&1&0&1
\end{bmatrix},
\: \quad
\bm{Z}_{ij}=  \begin{bmatrix}
1\\
1\\
1\\
1
\end{bmatrix},
$$
$$\bm{\beta}=(\mu,\pi_2,\tau_2,g_2).$$
 \item[(v)] Moreover, ${b}_{ij}=s_{ij}$ is the subject-specific random effect, and \\  $\bm{e}_{ij}=(e_{ij11}, \ldots ,e_{ijpm})^{T}$, is the random error vector. We assume an independent error structure to reduce the number of parameters and for ease of calculation in the maximum likelihood estimation. In our computations, we consider two scenarios:
 \begin{itemize}
     \item[] Scenario 1: error is skew-normal \begin{equation*}
\bm{e}_{ij} \overset{ind}{\sim} SN_{pm}(\bm{0},\sigma_{e}^2 \bm{I}_{pm},(\lambda_e,\ldots, 0)); \;\; {b}_{ij}  \overset{ind}{\sim} N(0,\sigma_s^2).
\end{equation*}
\item[] Scenario 2: random effect is skew-normal \begin{equation*}
\bm{e}_{ij} \overset{ind}{\sim} N_{pm}(\bm{0},\sigma_{e}^2 \bm{I}_{pm}); \;\; {b}_{ij}  \overset{ind}{\sim} SN(0,\sigma_s^2,\lambda_s).
\end{equation*}
 \end{itemize}%It is assumed 
\item[(vi)] The parameters of interest are,  $\bm{\theta}=(\bm{\beta}, \sigma_e^2,\sigma_s^2,\lambda_e,\lambda_s).$
\end{itemize}
For ease of the readers, we briefly define the univariate and multivariate skew normal distribution and their parameters used in the following sections (more details are given in \Cref{UN SN}). 
\begin{itemize}
    \item[(a)] Univariate skew-normal (SN) variate (\citet{azzalini1985}): If a random variable $W$ has the density function
    \begin{equation*}
    f(w;\lambda)=2\phi(w)\Phi(\lambda w);\;\; -\infty <w< \infty,
    \end{equation*}
    where $\phi$, $\Phi$ are the standard normal density and distribution function, respectively, then we say $W$ is a skew-normal variate with parameter $\lambda$, or, $W \sim SN(0,1,\lambda)$. 
  \item[(b)] Multivariate skew-normal (SN) variate (\citet{lachos2005}): An $n$ dimensional random vector $\bm{W}_1$ follows a skew-normal distribution with location vector $\bm{\mu} \in \mathbb{R}^{n}$, a dispersion matrix $\bm{\Sigma}$ (a $n \times n$ positive definite matrix) and skewness vector $\bm{\lambda} \in \mathbb{R}^{n}$, if its pdf is given by,
  \begin{equation*}
      f_{\bm{W}_1}(\bm{w}_1)=2\phi_n(\bm{w}_1|\bm{\mu},\bm{\Sigma})\Phi_1(\bm{\lambda}^{T}\bm{\Sigma}^{-1/2}(\bm{w}_1-\bm{\mu})),\;\; \bm{w}_1 \in \mathbb{R}^{n}.
  \end{equation*} 
  We denote it by $\bm{W}_1\sim SN_n(\bm{\mu},\bm{\Sigma},\bm{\lambda})$, further $\bm{W}_1$ can be written as $\bm{W}_1{=} \bm{\mu}+\bm{\Sigma}^{1/2}\bm{W}$, where, $\bm{W}$ is a standardized multivariate skew-normal vector. Throughout this article, skew-normal distributions are denoted by the letter SN, whereas normal distributions are identified by the letter N.
\end{itemize}

\begin{comment}
Under the above assumptions, following    
 \citet{lachos2005}, the marginal distribution  of $\bm{y}_{ij}$ is
\begin{align*}
 f_{\bm{y}_{ij}}(\bm{y}_{ij}|\bm{\theta} ) &= 2^2 \phi_{pm}(\bm{y}_{ij}|\bm{X}_{ij}\bm{\beta},\bm{\Sigma}_{ij})
 \\ &\times \Phi_2 \left((\bm{\mu}_{2ij}-\bm{\Gamma}_{ij}\bm{\mu}_{1ij})(\bm{y}_{ij}-\bm{X}_{ij}\bm{\beta})|\bm{0},\bm{I}_2+\bm{\Gamma}_{ij}{\Lambda}_{ij}\bm{\Gamma}_{ij}^{T}\right)\numberthis \label{marg lik}
\end{align*}
where, $\bm{\mu}_{1ij}$ is a vector of length $pm$, $\bm{\Sigma}_{ij}$ is a symmetric co-variance matrix of order $pm$, viz,
$$\bm{\mu}_{1ij}={\Lambda}_{ij}\bm{Z}_{ij}^{T}\sigma_e^{-2};\;\; \bm{\Sigma}_{ij}=\sigma_e^2 \bm{I}_{pm}+\bm{Z}_{ij}\sigma_s^{2}\bm{Z}_{ij}^{T};\;\; {\Lambda}_{ij}=(\sigma_s^{-2}+\bm{Z}_{ij}^{T}\sigma_e^{-2}\bm{Z}_{ij})^{-1} $$
\begin{align*}
    (\bm{\mu}_{2ij})_{2 \times pm} &= 
        \begin{pmatrix}
           (\lambda_e,0,\ldots, 0)\sigma_e^{-1} \\           
          (0,0,\ldots, 0)
           \end{pmatrix}
         \text{and}\; (\bm{\Gamma}_{ij})_{2 \times 1} =
          \begin{pmatrix}
        (\lambda_e,0,\ldots, 0)\sigma_e^{-1} \bm{Z}_{ij}\\           -\lambda_s \sigma_s^{-1}
           \end{pmatrix}
  \end{align*}
\end{comment}

\section{Maximum Likelihood Estimation}\label{mle sec}
As discussed in \citet{geertbook}, parameter estimation in the models as specified in \cref{mat form} is based on the marginal distribution of the response $\bm{y}_{ij}$ unless they are analyzed in the Bayesian framework. However, to compute the marginal distribution of $\bm{y}_{ij}$ we need to use complex Monte Carlo integration since it involves the skew-normal distribution. There is also no explicit method for directly maximizing the marginal likelihood, and one has to resort to numerical maximization (\citet{lachos2005}). 
%Also, there is no explicit solution available for direct maximization of the marginal likelihood, and one has to use the numerical maximization methods (\citet{lachos2005}). 

In this work, we use the EM algorithm (\citet{DEMP1977}, \citet{lachos2005}, \cite{lachos2010}), a generic iterative approach for maximum likelihood estimation in models with random effects and incomplete data. Every iteration of the EM algorithm has two steps: the Expectation Step (E-step) and the Maximization Step (M-step), which increases the likelihood function and typically converges to the local or global maximum of the likelihood function (\citet{wu1983}).% and the EM algorithm typically converges to  Although, it is robust to initial estimates and specifically it has 2 steps say one more paragraph....?

In this section, we discuss parameter estimation using the EM algorithm for two specific cases, (i) when the error is SN, and (ii) when the random effect is SN. The necessary results for the execution of the expectation step of the EM algorithm can be found in \Cref{appres}.
%To implement the expectation step of EM, results are given in \ref{appres}.
%we will use the results from \citet{lachos2005} given in \ref{appres}.
The corresponding maximization step is solved using the first-order Newton-Raphson equation.

\subsection{ML Estimation when Errors are SN}\label{error SN}
In the case of skew-normal errors, in \cref{mat form}, we assume
  $$ \bm{e}_{ij} \sim SN_{pm}(\textbf{0},\bm{V}_{1ij},\bm{\lambda}_{er}), \;\text{where,}\; \bm{V}_{1ij}=\sigma_{e}^2 \bm{I}_{pm}, \;\text{and} \; \bm{\lambda}_{er}=(\lambda_e,0,0,\ldots,0). $$ 
 Using \Cref{res 1} from the \Cref{appres}, we have
$$\bm{e}_{ij} \overset{d}{=}\bm{V}_{1ij}^{1/2}\bm{\delta}_{e}|U_{0ij}|+\bm{V}_{1ij}^{1/2}(\textbf{I}_{pm}-\bm{\delta}_{e}\bm{\delta}_{e}^{T})^{1/2}\textbf{U}_{1ij}, \; \;  $$
where, $\bm{\delta}_{e}=\frac{\bm{\lambda}_{er}}{\sqrt{1+\bm{\lambda}_{er}^{T}\bm{\lambda}_{er}}}$,  $U_{0ij} \sim N(0,1)$ independently of $\bm{U}_{1ij} \sim N_n(\bm{0},\bm{I}_n)$.
 Thus, \cref{mat form} can be written as
\begin{equation}
\bm{y}_{ij}=\bm{X}_{ij}\bm{\beta}+\bm{d}_{ij}t_{ij}+\bm{r}_{ij}, \label{mate eq}
\end{equation} where, $$\bm{d}_{ij}=\bm{V}_{1ij}^{1/2}\bm{\delta}_{e}, \;\; t_{ij}=|U_{0ij}|, \; t_{ij} \overset{iid}{\sim} HN_{1}(0,1),$$
 %$$\bm{r}_{ij}=\bm{Z}_{ij}{b}_{ij}+\bm{V}_{1ij}^{1/2}(\textbf{I}_{pm}-\bm{\delta}_{e}\bm{\delta}_{e}^{T})^{1/2}\textbf{U}_{1ij},$$
 and, $\bm{r}_{ij}=\bm{Z}_{ij}{b}_{ij}+\bm{V}_{1ij}^{1/2}(\textbf{I}_{pm}-\bm{\delta}_{e}\bm{\delta}_{e}^{T})^{1/2}\textbf{U}_{1ij},$ follows:
 $$\bm{r}_{ij} \overset{ind}{\sim} N_{pm}(\textbf{0},\bm{V}_{ij}),\; \text{for}\; \bm{V}_{ij}=\sigma_s^2 \bm{Z}_{ij}\bm{Z}_{ij}^T+ \sigma_e^2 \bm{R},\;\; \bm{R}=\textbf{I}_{pm}-\bm{\delta}_{e}\bm{\delta}_{e}^{T}.$$ 
 Moreover, for $i = 1(1)s$ and $j = 1(1)n_{i}$, $t_{ij} \perp \bm{r}_{ij}$. Since $\bm{r}_{ij}$ has a zero mean, it can be used in residual analyses to assess the adequacy of the model. The right side of \cref{mate eq} has a mean $\sigma_e \bm{\delta}_e\sqrt{2/ \pi} $, which is used to correct the intercept in the fixed effects. Under the above setup, the conditional model where conditioning refers to $t_{ij}$ takes the form
$$ \bm{y}_{ij}|t_{ij} \sim N_{pm}(\bm{\mu}_{ij}^{*},\bm{V}_{ij}),\; \bm{\mu}_{ij}^{*}=\bm{X}_{ij}\bm{\beta}+\bm{d}_{ij}t_{ij}. $$ 

The complete data log-likelihood can then be expressed as
\begin{align*}
l_{c}(\bm{\theta}_{c})&=\log(\prod_{i=1}^{s}\prod_{j=1}^{n_{i}}f(\bm{y}_{ij},t_{ij}))=\sum_{i=1}^{s}\sum_{j=1}^{n_{i}}\left(\log f(\bm{y}_{ij}|t_{ij}) + \log f(t_{ij})\right)\\
& \propto -\frac{1}{2} \sum_{i=1}^{s}\sum_{j=1}^{n_{i}}\left(\log |\bm{V}_{ij}| + (\bm{y}_{ij}-\bm{\mu}_{ij}^{*})^{T}{\bm{V}_{ij}}^{-1}(\bm{y}_{ij}-\bm{\mu}_{ij}^{*})+t_{ij}^2\right).\numberthis \label{comp likeli}
\end{align*}
%$$\bm{\theta}_{c}=(\bm{\beta}^{T},\lambda, \sigma_{e}^{2},\sigma_{s}^{2})$$
The E-step consists of calculating the expected value of the complete data log-likelihood given the observed data and current parameter estimates. 

For fixed i, j, we have \begin{align*}
Q_{ij}(\bm{\theta}_{c}|\bm{\theta}_{c}^{(r)})&= E_{t_{ij}|\bm{y}_{ij},\bm{\theta}_{c}^{(r)}}[l_{c}(\bm{\theta}_{c})]\numberthis \label{q fn}
= \int l_{c}(\bm{\theta}_{c}) f(t_{ij}|\bm{y}_{ij},\bm{\theta}_{c}^{(r)}) dt_{ij}.
\end{align*}

Using $l_{c}(\bm{\theta}_{c})$ from \cref{comp likeli} in \cref{q fn}, the expression for $Q_{ij}$ is as follows,%\cref{comp likeli} in \cref{q fn} we get the following,
\begin{align*}
Q_{ij}(\bm{\theta}_{c}|\bm{\theta}_{c}^{(r)})&= -\frac{1}{2}[\log |\bm{V}_{ij}|+E_{t_{ij}|\bm{y}_{ij},\bm{\theta}_{c}^{(r)}}(t_{ij}^2)
\\&+E_{t_{ij}|\bm{y}_{ij},\bm{\theta}_{c}^{(r)}}(\bm{y}_{ij}-\bm{X}_{ij}\bm{\beta}-\bm{d}_{ij}t_{ij})^{T}{\bm{V}_{ij}}^{-1}(\bm{y}_{ij}-\bm{X}_{ij}\bm{\beta}-\bm{d}_{ij}t_{ij})] 
\\ &=-\frac{1}{2}[\log |\bm{V}_{ij}|+T_{ij}^{02}+\bm{d}_{ij}^T {\bm{V}_{ij}}^{-1}\bm{d}_{ij} T_{ij}^{02}
\\&+(\bm{y}_{ij}-\bm{X}_{ij}\bm{\beta})^{T}{\bm{V}_{ij}}^{-1}(\bm{y}_{ij}-\bm{X}_{ij}\bm{\beta}-2\bm{d}_{ij}T_{ij}^{01})], %\numberthis \label{qij}
\end{align*}
where, $T_{ij}^{01}$, and $T_{ij}^{02}$ are obtained as follows using Lemmas \ref{res 2}, and \ref{res 3} from the \Cref{appres};% $T_{ij}^{01}=E[t_{ij}|\bm{y}_{ij},\bm{\theta}_{c}^{(r)}]$, $T_{ij}^{02}=E[t_{ij}^2|\bm{y}_{ij},\bm{\theta}_{c}^{(r)}].$
$$T_{ij}^{01} = E[t_{ij}|\bm{y}_{ij},\bm{\theta}_{c}^{(r)}]=\eta_{ij}+\frac{\phi(\frac{\eta_{ij}}{\zeta_{ij}})}{\Phi_{1}(\frac{\eta_{ij}}{\zeta_{ij}})}\zeta_{ij}.$$ 
$$T_{ij}^{02}= E[t_{ij}^{2}|\bm{y}_{ij},\bm{\theta}_{c}^{(r)}]=\eta_{ij}^{2}+\zeta_{ij}^{2}+\frac{\phi(\frac{\eta_{ij}}{\zeta_{ij}})}{\Phi_{1}(\frac{\eta_{ij}}{\zeta_{ij}})}\eta_{ij}\zeta_{ij},$$
$$\text{here,}\;\eta_{ij}= \frac{\bm{d}_{ij}^{T}\bm{V}_{ij}^{-1}(\bm{y}_{ij}-\bm{X}_{ij}\bm{\beta})}{1+\bm{d}_{ij}^{T}\bm{V}_{ij}^{-1}\bm{d}_{ij}}\;\; \text{and} \; \zeta_{ij}^{2}=\frac{1}{1+\bm{d}_{ij}^{T}\bm{V}_{ij}^{-1}\bm{d}_{ij}}.$$
%As observations from each subject are assumed to be independent, T
Considering subjects are independent, the E-step yields,
\begin{equation}\label{final q}
  Q(\bm{\theta}_{c}|\bm{\theta}_{c}^{(r)})= \sum_{i=1}^{s}\sum_{j=1}^{n_{i}} Q_{ij}(\bm{\theta}_{c}|\bm{\theta}_{c}^{(r)}). 
\end{equation}
In the second step of the algorithm, \textit{i.e.,} the M-step, our task is to maximize $Q(\bm{\theta}_{c}|\bm{\theta}_{c}^{(r)})$, we do this in two parts. Here the maximization of $Q-$function corresponding to the fixed effects is straightforward and closed-form expressions are available. However, for variance components, maximization of $Q-$function has to be done numerically using the first-order Newton-Raphson equation. 
In the M-step as a first step, we find the $(r+1)^{th}$ estimate of ${\bm{\beta}}$, $\bm{\beta}^{(r+1)}$, by maximizing
\begin{comment}The closed-form expression is available for $\bm{\beta}$. To get the estimates of $(\sigma_e^2, \sigma_s^2, \lambda)$ Newton-Raphson method is used to solve the first order derivative of $Q(\bm{\theta}_{c}|\bm{\theta}_{c}^{(r)})$. 
\end{comment}
\begin{align*}
Q(\bm{\beta})&=\sum_{i=1}^{s}\sum_{j=1}^{n_{i}} -\frac{1}{2}[(\bm{y}_{ij}-\bm{X}_{ij}\bm{\beta})^{T}{\bm{V}_{ij}}^{-1}(\bm{y}_{ij}-\bm{X}_{ij}\bm{\beta}-2\bm{d}_{ij}T_{ij}^{01})],
\end{align*}
which yields
\begin{align*}
\bm{\beta}^{(r+1)}&=(\sum_{i=1}^{s}\sum_{j=1}^{n_{i}}\bm{X}_{ij}^{T}{\bm{V}_{ij}}^{-1(r)} \bm{X}_{ij})^{-1} \sum_{i=1}^{s}\sum_{j=1}^{n_{i}}\bm{X}_{ij}^{T}{\bm{V}_{ij}}^{-1(r)}(\bm{y}_{ij}-\bm{d}_{ij}T_{ij}^{01(r)}).
\end{align*}
We use these estimated $\bm{\beta}^{(r+1)}$ in the second step to obtain the variance components. The three components $(\sigma_e^{2(r+1)}, \sigma_s^{2(r+1)}, \lambda_e^{(r+1)})$ are found numerically by solving the first order derivatives of the $Q-$ function (\cref{final q}), using the Newton-Raphson method. Suppose ${\bm{\xi}}=({\xi}_{[1]},{\xi}_{[2]},{\xi}_{[3]})=(\sigma_e^2,\sigma_s^2,\lambda_e)$, then,
\begin{align*}
\frac{\partial Q}{\partial{\xi}_{[v_0]}}&=\sum_{i=1}^{s}\sum_{j=1}^{n_{i}}-\frac{1}{2}[\text{tr}(\bm{V}_{ij}^{-1}\bm{V}_{ij{\xi}_{[v_0]}})+\bm{d}_{ij}^T {\bm{V}^{*}_{ij{\xi}_{[v_0]}}}\bm{d}_{ij} T_{ij}^{02}+2\bm{d}_{ij}^T\bm{V}_{ij}^{-1}{\bm{d}_{ij{\xi}_{[v_0]}}} T_{ij}^{02}\\
&+(\bm{y}_{ij}-\bm{X}_{ij}\bm{\beta}^{(r+1)})^{T}{\bm{V}^{*}_{ij{\xi}_{[v_0]}}}(\bm{y}_{ij}-\bm{X}_{ij}\bm{\beta}^{(r+1)}-2\bm{d}_{ij}T_{ij}^{01})\\
&-2(\bm{y}_{ij}-\bm{X}_{ij}\bm{\beta}^{(r+1)})^{T}\bm{V}_{ij}^{-1}{\bm{d}_{ij{\xi}_{[v_0]}}}T_{ij}^{01}],\;\; v_0 =1, 2, 3,
\end{align*}
where, \begin{align*}\bm{V}_{ij\bm{\xi}}&=(\frac{\partial \bm{V}_{ij}}{\partial\sigma_e^2 },\frac{\partial \bm{V}_{ij}}{\partial\sigma_s^2 },\frac{\partial \bm{V}_{ij}}{\partial\lambda_e })=(\bm{R},\bm{Z}_{ij}\bm{Z}_{ij}^{T},\sigma_e^2 \bm{R}_{l}),\\
\bm{d}_{ij\bm{\xi}}&=(\frac{\partial \bm{d}_{ij}}{\partial\sigma_e^2 },\frac{\partial \bm{d}_{ij}}{\partial\sigma_s^2 },\frac{\partial \bm{d}_{ij}}{\partial\lambda_e })=(\frac{\bm{\delta}_{e}}{2 \sqrt{\sigma_e^2}},\bm{0},\sigma_e \bm{\delta}_{el}),\\
\bm{V}^{*}_{ij\bm{\xi}}&=\frac{\partial \bm{V}^{-1}_{ij}}{\partial\bm{\xi} }=-\bm{V}_{ij}^{-1}\bm{V}_{ij{\bm{\xi}}}\bm{V}_{ij}^{-1},\\ \text{and}\;
\bm{\delta}_{el}&=\frac{\partial \bm{\delta}_e}{\partial \lambda_e}=\frac{(1,0,\ldots,0)}{(1+\lambda_e^2)^{3/2}},\;\;
\bm{R}_l=\frac{\partial \bm{R}}{\partial \lambda_e}=-\bm{\delta}_e \bm{\delta}_{el}^{T}-\bm{\delta}_{el} \bm{\delta}_{e}^{T}.
\end{align*}

For the parameter vector $\bm{\xi}=(\sigma_e^2,\sigma_s^2,\lambda_e)$, the updated estimates at $(r+1)^{\text{th}}$ iteration are then given by
\begin{equation*}
\bm{\xi}^{(r+1)}=\bm{\xi}^{(r)}- \bm{H}^{-1(r)}\frac{\partial Q}{\partial \bm{\xi}}\Bigr|_{\bm{\xi}=\bm{\xi}^{(r)}}, \;\text{where,}\; \bm{H}=\frac{\partial^2 Q}{\partial \bm{\xi} \partial \bm{\xi}^{'}}.
\end{equation*}
Detailed computations of the $\bm{H}$ matrix are provided in \Cref{rerr}. We define the convergence criterion as the difference between estimated values at $(r+1)^{\text{th}}$ iteration and $(r)^{\text{th}}$ iteration being less than $5\times 10^{-3}$. The E and M-steps are iterated until convergence. 

\subsection{ML Estimation when Random Effects are SN}\label{random SN}
For skew-normal random effects, \textit{i.e.,} when ${b}_{ij} \sim SN(0,\sigma_s^2,\lambda_s)$, applying \Cref{res 1} of \Cref{appres}, we can express ${b}_{ij}$ in \cref{mat form} as
%${b}_{ij} \sim SN(0,\sigma_s^2,\lambda_s)$
$${b}_{ij} \overset{d}{=}\sigma_s {\delta}_{b}|U_{0ij}|+\sigma_s(1-\delta_{b}^2)^{1/2}{U}_{1ij}, \; \; $$
where, ${\delta}_{b}=\frac{{\lambda_s}}{\sqrt{1+\lambda_s^2}}$, $U_{0ij}, \; \text{and}\; U_{1ij}$ are independent standard normals. As a result, \cref{mat form} becomes
\begin{equation}\label{mate eq1}
\bm{y}_{ij}=\bm{X}_{ij}\bm{\beta}+\bm{d}_{ij}t_{ij}+\bm{r}_{ij},
\end{equation} 
where, $$\bm{d}_{ij}=\bm{Z}_{ij}\sigma_s \delta_b; \; t_{ij}=|U_{0ij}|,\  \text{and}\; t_{ij} \overset{iid}{\sim} HN_{1}(0,1)$$
%$$t_{ij}=|U_{0ij}|,\;\; t_{ij} \overset{iid}{\sim} HN_{1}(0,1),$$
 $$\text{For}\; \bm{r}_{ij}=\bm{Z}_{ij}\sigma_s (1-\delta_b^2)^{1/2}\bm{U}_{1ij}+\bm{e}_{ij},   $$
we see that,
 $$\bm{r}_{ij} \overset{ind}{\sim} N_{pm}(\textbf{0},\bm{V}_{ij}), \; \text{where}\;
 \bm{V}_{ij}=\sigma_s^2 \bm{Z}_{ij}R\bm{Z}_{ij}^T+ \sigma_e^2 \bm{I}_{pm},\;\text{and}\; R=1-\delta_b^2.$$
Further, $ t_{ij} \perp \bm{r}_{ij} $; $i = 1(1)s$ and $j = 1(1)n_{i}$. %Note that the mean of $\bm{r}_{ij}$ is zero, so it can be used, for example, in residual analysis to check model adequacy. /cref[mate eq] has a mean $/bm[Z]_[ij]/sigma_s /delta_b$ that will be used to correct the intercept in the fixed effects so that their interpretation is the same as that of a traditional LMM (population average) (\citet{lachos2005}).
\Cref{mate eq1} has a mean $\bm{Z}_{ij}\sigma_s \delta_b$ that is used to correct the intercept in the fixed effects. Under the above setting, the conditional model where conditioning refers to $t_{ij}$ takes the form,
$$ \bm{y}_{ij}|t_{ij} \sim N_{pm}(\bm{\mu}_{ij}^{*},\bm{V}_{ij}),\; \bm{\mu}_{ij}^{*}=\bm{X}_{ij}\bm{\beta}+\bm{d}_{ij}t_{ij}.$$ 

As in \Cref{error SN}, the complete data log-likelihood can be expressed as follows:
\begin{align*}
l_{c}(\bm{\theta}_{c})&\propto -\frac{1}{2} \sum_{i=1}^{s}\sum_{j=1}^{n_{i}}\left(\log |\bm{V}_{ij}| + (\bm{y}_{ij}-\bm{\mu}_{ij}^{*})^{T}{\bm{V}_{ij}}^{-1}(\bm{y}_{ij}-\bm{\mu}_{ij}^{*})+t_{ij}^2\right).
\end{align*}
%$$\bm{\theta}_{c}=(\bm{\beta}^{T},\lambda, \sigma_{e}^{2},\sigma_{s}^{2})$$
The E-step, as discussed for the skewed random error case, consists of calculating the expected value of complete data log-likelihood given the observed data and current parameter estimates. 
\begin{comment}
For fixed i, j, \begin{align*}
Q_{ij}(\bm{\theta}_{c}|\bm{\theta}_{c}^{(r)})&= E_{t_{ij}|\bm{y}_{ij},\bm{\theta}_{c}^{(r)}}[l_{c}(\bm{\theta}_{c})]\numberthis \label{q fn11}\\
&= \int l_{c}(\bm{\theta}_{c}) f(t_{ij}|\bm{y}_{ij},\bm{\theta}_{c}^{(r)}) dt_{ij},
\end{align*}

Using \cref{res 2}, \cref{res 3} and \cref{comp likeli11} in \cref{q fn11} we get the following,
\begin{align*}
Q_{ij}(\bm{\theta}_{c}|\bm{\theta}_{c}^{(r)})&= -\frac{1}{2}[\log |\bm{V}_{ij}|+E_{t_{ij}|\bm{y}_{ij},\bm{\theta}_{c}^{(r)}}(t_{ij}^2)
\\&+E_{t_{ij}|\bm{y}_{ij},\bm{\theta}_{c}^{(r)}}(\bm{y}_{ij}-\bm{X}_{ij}\bm{\beta}-\bm{d}_{ij}t_{ij})^{T}{\bm{V}_{ij}}^{-1}(\bm{y}_{ij}-\bm{X}_{ij}\bm{\beta}-\bm{d}_{ij}t_{ij})] 
\end{align*}
\end{comment}
\begin{align*}
Q_{ij}(\bm{\theta}_{c}|\bm{\theta}_{c}^{(r)})&=-\frac{1}{2}[\log |\bm{V}_{ij}|+T_{ij}^{02}+\bm{d}_{ij}^T {\bm{V}_{ij}}^{-1}\bm{d}_{ij} T_{ij}^{02}
\\&+(\bm{y}_{ij}-\bm{X}_{ij}\bm{\beta})^{T}{\bm{V}_{ij}}^{-1}(\bm{y}_{ij}-\bm{X}_{ij}\bm{\beta}-2\bm{d}_{ij}T_{ij}^{01})]%\numberthis \label{comp likeli11}
\end{align*}
where, $T_{ij}^{01}=E[t_{ij}|\bm{y}_{ij},\bm{\theta}_{c}^{(r)}]$, $T_{ij}^{02}=E[t_{ij}^2|\bm{y}_{ij},\bm{\theta}_{c}^{(r)}].$ The E-step for all the subjects yields,
\begin{equation}
  Q(\bm{\theta}_{c}|\bm{\theta}_{c}^{(r)})= \sum_{i=1}^{s}\sum_{j=1}^{n_{i}} Q_{ij}(\bm{\theta}_{c}|\bm{\theta}_{c}^{(r)}). \label{qfn1} 
\end{equation}
Similar to M-step in \Cref{error SN}, we maximize $Q(\bm{\theta}_{c}|\bm{\theta}_{c}^{(r)})$ in two parts. In the first part we find $\bm{\beta}^{(r+1)}$ by maximising \cref{qfn1} with respect to $\bm{\beta}$,
\begin{align*}
\bm{\beta}^{(r+1)}&=(\sum_{i=1}^{s}\sum_{j=1}^{n_{i}}\bm{X}_{ij}^{T}{\bm{V}_{ij}}^{-1(r)} \bm{X}_{ij})^{-1} \sum_{i=1}^{s}\sum_{j=1}^{n_{i}}\bm{X}_{ij}^{T}{\bm{V}_{ij}}^{-1(r)}(\bm{y}_{ij}-\bm{d}_{ij}T_{ij}^{01(r)}).
\end{align*}
While in the second part, we numerically solve for $(\sigma_e^{2(r+1)}, \sigma_s^{2(r+1)}, \lambda_s^{(r+1)})$ using the Newton-Raphson method. Based on \Cref{error SN}, ${\bm{\xi}}=({\xi}_{[1]},{\xi}_{[2]},{\xi}_{[3]})=(\sigma_e^2,\sigma_s^2,\lambda_s)$, then 
\begin{align*}
\frac{\partial Q}{\partial{\xi}_{[v_0]}}&=\sum_{i=1}^{s}\sum_{j=1}^{n_{i}}-\frac{1}{2}[\text{tr}(\bm{V}_{ij}^{-1}\bm{V}_{ij{\xi}_{[v_0]}})+\bm{d}_{ij}^T {\bm{V}^{*}_{ij{\xi}_{[v_0]}}}\bm{d}_{ij} T_{ij}^{02}+2\bm{d}_{ij}^T\bm{V}_{ij}^{-1}{\bm{d}_{ij{\xi}_{[v_0]}}} T_{ij}^{02}\\
&+(\bm{y}_{ij}-\bm{X}_{ij}\bm{\beta})^{T}{\bm{V}^{*}_{ij{\xi}_{[v_0]}}}(\bm{y}_{ij}-\bm{X}_{ij}\bm{\beta}-2\bm{d}_{ij}T_{ij}^{01})
\\&-2(\bm{y}_{ij}-\bm{X}_{ij}\bm{\beta})^{T}\bm{V}_{ij}^{-1}{\bm{d}_{ij{\xi}_{[v_0]}}}T_{ij}^{01}], v_0 =1, 2, 3,
\end{align*}
where, \begin{align*}\bm{V}_{ij\bm{\xi}}&=(\frac{\partial \bm{V}_{ij}}{\partial\sigma_e^2 },\frac{\partial \bm{V}_{ij}}{\partial\sigma_s^2 },\frac{\partial \bm{V}_{ij}}{\partial\lambda_s })=(\bm{I}_{pm},R\bm{Z}_{ij}\bm{Z}_{ij}^{T},\sigma_s^2 R_{l}\bm{Z}_{ij}\bm{Z}_{ij}^{T}),\\
\bm{d}_{ij\bm{\xi}}&=(\frac{\partial \bm{d}_{ij}}{\partial\sigma_e^2 },\frac{\partial \bm{d}_{ij}}{\partial\sigma_s^2 },\frac{\partial \bm{d}_{ij}}{\partial\lambda_s })=(\bm{0},\frac{\bm{Z}_{ij}\delta_b}{2 \sqrt{\sigma_s^2}},\bm{Z}_{ij}\sigma_s\delta_{bl}),\\
\bm{V}^{*}_{ij\bm{\xi}}&=\frac{\partial \bm{V}^{-1}_{ij}}{\partial\bm{{\xi}} }=-\bm{V}_{ij}^{-1}\bm{V}_{ij{\bm{\xi}}}\bm{V}_{ij}^{-1},\\
\delta_{bl}&=\frac{\partial \delta_b}{\partial \lambda_s}=\frac{1}{(1+\lambda_s^2)^{3/2}},\;\; \text{and}\;
R_l=\frac{\partial R}{\partial \lambda_s}=-2\delta_b\delta_{bl}.
\end{align*}

For the parameter vector $\bm{\xi}=(\sigma_e^2,\sigma_s^2,\lambda_s)$, the updated estimates at the $(r+1)^{\text{th}}$ iteration are given by
\begin{equation*}
\bm{\xi}^{(r+1)}=\bm{\xi}^{(r)}- \bm{H}^{-1(r)}\frac{\partial Q}{\partial \bm{\xi}}\Bigr|_{\bm{\xi}=\bm{\xi}^{(r)}}, \bm{H}=\frac{\partial^2 Q}{\partial \bm{\xi} \partial \bm{\xi}^{'}}.
\end{equation*}

The elements of the $\bm{H}$ matrix are given in \Cref{rskew}. The E and M-steps are iterated till convergence is achieved.

\section{Simulation Studies}\label{sim study}
To assess the performance of the proposed estimators, we present simulation studies for the two cases, (i) errors are skew-normal and (ii) random effects are skew-normal. For data generation in both cases, we assume a crossover trial with three treatment sequences $\{ABC, BCA, CAB\}$ in three periods. Two simulation scenarios with 30 and
50 subjects assigned, respectively to each treatment sequence are considered. In each period, four response variates are measured. The model is represented as,
%The model in regression setting:
\begin{align*}
y_{i j tk}&=\beta_{0}+\beta_{p_2} \text{Per}_2 +\beta_{p_3}  \text{Per}_3+\beta_{\tau_2} \text {Trt}_{2}+\beta_{\tau_3} \text {Trt}_{3}+\beta_{g_2} \text {Gene}_{2}+\beta_{g_3} \text {Gene}_{3}\\&+\beta_{g_4}\text {Gene}_{4}+\beta_1 w_{ij}+s_{i j} 
+e_{ijtk};\;i, t=1, 2, 3; \;j=1 (1) n_i; \;k=1 (1) 4, %\numberthis \label{model in regressionn}
\end{align*}
where $y_{i jtk}$ denotes the $k^{\text{th}}$ response value from the $j^{\text{th}}$ subject in the $t^{\text{th}}$ period of the $i^{\text{th}}$ sequence; $\beta_{0}$: intercept; Per$_2$, Per$_3$: indicator variables corresponding to time/period effects, %where Per$_{i}=1$ for the $i^{\text{th}}$ period and $0$ otherwise for $i=2, 3$; 
Trt$_{2}$, Trt$_{3}$:  indicator variables corresponding to the treatment effects, % where Trt$_{i}=1$ for the $i^{\text{th}}$ treatment and $0$ otherwise for $i=2, 3$ 
and Gene$_{2}$, Gene$_{3}$, Gene$_{4}$: indicator variables corresponding to a subject's gene expression level. Following are the indicated variables with respect to the period, treatment, and gene effects for $u=2, 3$ and $v=2, 3, 4$, %where Gene$_{i}=1$ for the $i^{\text{th}}$ gene expression and $0$ otherwise for $i=2, 3, 4$;
\begin{equation*}
 \text{Per}_{u}/\text{Trt}_u =\begin{cases}
    1, & \text{if $u^{\text{th}}$ period/treatment}\\
    0, & \text{otherwise}
  \end{cases} 
\end{equation*}
\begin{equation*}
 \text{Gene}_{v}=\begin{cases}
    1, & \text{if $v^{\text{th}}$ gene expression level}\\
    0, & \text{otherwise}
  \end{cases} 
\end{equation*}
$w_{ij}$ represents an individual-level covariate taking values in $\{0, 1, 2\}$. In scenario 1, when there are 30 subjects in each sequence, $w_{ij}$ takes the value 0 for subjects 1 to 10, 1 for subjects 11 to 20, and 2 for subjects 21 to 30. Considering scenario 2, when $n_i=50$ for each sequence, $w_{ij}$ is 0 for subjects 1 to 18, 1 for subjects 19 to 34, and 2 for subjects 35 to subject 50. %with $w_ij=0$ for the first 10 subjects, $w_{ij}=1$ for next 10 subject and $w_{ij}=2$ for the remaining 10 subject of each sequence. $w_ij=0, if j<=no[I]/3$ from uniform random variate with parameter $(i, i+2)$ for each sequence $i= 1, 2, 3$. 
Also, $s_{i j}$ and $e_{i j tk}$ are the subject-specific random effect and the random error terms, respectively.
%\[
%s_{i k} \sim N\left(0, \sigma_{s}^{2}\right), \quad e_{i j kl} \sim N\left(0, \sigma_{e}^{2}\right);\; \; s_{ik} \perp e_{ijkl}
%\]
In matrix notations,
\begin{equation*}%\label{eq:sim model}
\bm{y}_{ij}=\bm{ X}_{ij}\bm{\beta}+\bm{Z}_{ij}{b}_{ij}+\bm{e}_{ij};\;\; i=1, 2, 3, \; \;j=1, 2, \ldots, n_i,
\end{equation*}
where, $\bm{X}_{ij}$ and $\bm{Z}_{ij}$ matrices are as described in \cref{mat form}. %In sequence, $i= 1, 2, 3$, the individual-level covariate $w_{ij}$ is generated from uniform random variate with parameter $(i, i+2)$. 
The true values of the components of $\bm{\beta}$ are given in the first column of Tables \ref{table simulated data} and \ref{table simulated data1}, respectively. For our simulations, we take the following true values for the variance components:
\begin{itemize}
    \item[(i)] For the case where the errors are skew-normal, $$ b_{ij}\sim N(0, 0.64), \bm{e}_{ij} \sim SN_{pm} (\bm{0}, 2\bm{I}_{pm}, (3,0,\ldots,0)).$$ Hence, $\sigma_e^2 = 2, \sigma_s^2 = 0.64,$ and $\lambda_e= 3.$
    \item[(ii)] For the skewed random effect case, $$ b_{ij}\sim SN(0, 3, 4), \bm{e}_{ij} \sim N_{pm} (\bm{0}, 0.72\bm{I}_{pm}).$$
    Thus, $\sigma_e^2 = 0.72, \sigma_s^2 = 3,$ and $\lambda_s = 4.$
\end{itemize}

Two hundred Monte Carlo data sets were generated for each of the above parameter settings. The model fitting results under parameter settings (i) and (ii) are compared with the case where we assume both random effects and random errors to be normally distributed. The Akaike Information Criterion (AIC) (\citet{glos1994}) was applied to select the best-fitting model. 
%We have used the Akaike Information Criterion (AIC) to select the best-fitted model as in \citet{lachos2005}. 
%We also compared the fitting of the proposed model with that described in \citet{schu2020}, where they assume that the random effect is a scale mixture of skew-normal and that the random error is a scale mixture of normal distribution. We chose to compare only the model fitted under parameter setting (ii), since the \citet{schu2020} discusses only the random effect as skew-normal.
%Furthermore, the model fitting under parameter setting (ii) was also compared with the methodology described in \citet{schu2020}, where they consider the random effect to be a scale mixture of skew-normal and random error to scale mixture of normal. for the  smsn.lmm function of skewlmm package
%With each simulated data set in setting (i), the model (\cref{model in regressionn}) has been fitted under the assumption that (a) random error is skew-normal and (b) random error is normal. Similarly, for every simulated data set in setting (ii), the model (\cref{model in regressionn}) has been fitted assuming (a) random effect is skew-normal and (b) random effect is normal. Akaike Information Criterion (AIC) is used to select the best-fitted model (\citet{lachos2010}).

After the estimation of all of the model parameters, the fitted distribution of $\bm{r}_{ij}$ given in \cref{mate eq}, \cref{mate eq1} can be plotted along with the data scatter in order to assess the model's adequacy. In an alternative approach to evaluating model fitting, Mahalanobis-type distances may be used, which were first introduced by \citet{healy1968} for multivariate normal distribution distances. Let $\bm{y}=(\bm{y}_1,\ldots, \bm{y}_m)$ where the $i^{\text{th}}$ component, $\bm{y}_i$, is sampled from $SN_n(\bm{\mu},\bm{\Sigma},\bm{\lambda})$, the Mahalanobis-type distances are defined as, $$d_i= (\bm{y}_i-\bm{\hat{\mu}})^{T}\bm{\hat{\Sigma}}^{-1}(\bm{y}_i-\bm{\hat{\mu}}),\;\; i= 1, 2, \ldots, m,$$
whose approximate reference distribution is $\chi^2_n$ (\citet{azzalini2014}. From these $d_i$'s, we obtain QQ-plots. Plotting nominal probability values against the theoretical cumulative probabilities of the observed Mahalanobis distances. These plots are also called a Healy-type plot (\citet{schu2020}). An appropriately fitted model should produce a straight line with a unit slope through the origin in a Healy-type plot. %mention rij also....

Tables \ref{table simulated data} and \ref{table simulated data1} show the average simulation results in terms of parameter estimates, standard errors (SEs), and absolute bias. The SEs are estimated using the Hessian matrix, and the average absolute bias for the true value $\beta_u$ is computed  as $\frac{\sum_{w=1}^{200} |\hat{\beta}_{uw}-\beta_u|}{200 }$, where $\hat{\beta}_{uw}$ is the $u^\text{th}$ component of $\hat{\bm{\beta}}$ for the $w^\text{th}$ simulation. %Further, AIC is computed using $-2* Q$-fn in the case of EM- algorithm as in \citet{glos1994}.%Whereas, the empirical coverage probability (ECP) was taken to be the proportion of the times the $95\%$ asymptotic confidence interval of $\hat{\beta}_{uw}$ contained the true parameter value $\beta_u$ in $500$ simulations.  
%To obtain the SEs, one hundred imputations with a burn-in of $20\%$ were used. 

\begin{table}[ht]
\scriptsize
\centering
\caption{Simulation study: results based on 200 Monte Carlo samples with a different number of subjects. Maximum likelihood estimates, SEs, and absolute bias for the cases $ b_{ij}\sim N(0, 0.6),\; \bm{e}_{ij} \sim SN_{pm} (\bm{0}, 2\bm{I}_{pm}, (3,0,\ldots,0))$ and when both $\bm{e}_{ij}$, $b_{ij}$ are N. True parameter values are given in parentheses.
}
\label{table simulated data}
\begin{tabular}{lllllll} \hline
& \multicolumn{3}{c}{If $\bm{e}_{ij}$ is SN, $b_{ij}$ is N} &\multicolumn{3}{c}{Both $\bm{e}_{ij}$ , $b_{ij}$ are N}       \\  \hline
Parameter  & Estimate        & SE       & $\left|\text{Bias}\right|$        & Estimate & SE & $\left|\text{Bias}\right|$    \\ \hline
\multicolumn{7}{c}{$n_i$=30 for each sequence}                                              \\  \hline
$\beta_0$(2.1)                                & 2.0769          & 0.0505            & 0.1459       & 1.5482   & 0.1846      & 0.5225 \\
$\beta_{p_2}$(2.4)                                     & 2.3925          & 0.0767            & 0.0811       & 2.1271   & 0.1036      & 0.2729 \\
$\beta_{p_3}$(1.1)                                   & 1.0826          & 0.0797            & 0.0807       & 0.8172   & 0.1036      & 0.2828 \\
$\beta_{\tau_2}$(0.9)                                   & 0.8999          & 0.0793            & 0.0770       & 0.9026   & 0.1036      & 0.0810 \\
$\beta_{\tau_3}$(2.1)                                   & 2.1118          & 0.0533            & 0.0737       & 2.1130   & 0.1036      & 0.0793 \\
$\beta_{g_2}$(1.5)                                        & 1.4750          & 0.0820            & 0.0882       & 1.1211   & 0.1197      & 0.3789 \\
$\beta_{g_3}$(2.0)                                        & 1.9880          & 0.0928            & 0.0864       & 1.6341   & 0.1197      & 0.3659 \\
$\beta_{g_4}$(3.4)                                        & 3.3853          & 0.0661            & 0.0814       & 3.0314   & 0.1197      & 0.3686 \\
$\beta_{1}$(1.8)                                   & 1.7959          & 0.0501            & 0.1049       & 1.7966   & 0.1136      & 0.1045 \\
$\sigma_e^2$(2.0) & 1.9774          & 0.1182            & 0.0746       & 1.9343   & 1.3904      & 0.0914 \\
$\sigma_s^2$(0.6) & 0.6216          & 0.0461            & 0.0922       & 0.6166   & 0.7821      & 0.0939 \\
$\lambda_e$(3.0)                      & 3.2824          & 0.2384            & 0.8926       &  -        &     -        &   -     \\ \hline
\multicolumn{7}{c}{$n_i$=50 for each sequence}                                              \\ \hline
$\beta_0$(2.1)                                & 2.0690          & 0.0322            & 0.1120       & 1.5314   & 0.1410      & 0.5391 \\
$\beta_{p_2}$(2.4)                                     & 2.4009          & 0.0603            & 0.0642       & 2.1345   & 0.0806      & 0.2655 \\
$\beta_{p_3}$(1.1)                                   & 1.0919          & 0.0634            & 0.0624       & 0.8254   & 0.0806      & 0.2746 \\
$\beta_{\tau_2}$(0.9)                                   & 0.9039          & 0.0619            & 0.0670       & 0.9058   & 0.0806      & 0.0666 \\
$\beta_{\tau_3}$(2.1)                                   & 2.1062          & 0.0413            & 0.0621       & 2.1058   & 0.0806      & 0.0648 \\
$\beta_{g_2}$(1.5)                                        & 1.4886          & 0.0651            & 0.0668       & 1.1333   & 0.0930      & 0.3667 \\
$\beta_{g_3}$(2.0)                                        & 2.0075          & 0.0728            & 0.0802       & 1.6522   & 0.0930      & 0.3478 \\
$\beta_{g_4}$(3.4)                                        & 3.4049          & 0.0507            & 0.0726       & 3.0497   & 0.0930      & 0.3503 \\
$\beta_{1}$(1.8)                                   & 1.8011          & 0.0414            & 0.0708       & 1.8015   & 0.0877      & 0.0695 \\
$\sigma_e^2$(2.0) & 1.9906          & 0.0867            & 0.0580       & 1.9471   & 1.3952      & 0.0720 \\
$\sigma_s^2$(0.6) & 0.6260          & 0.0388            & 0.0726       & 0.6221   & 0.7867      & 0.0740 \\
$\lambda_e$(3.0)                      & 3.4737          & 0.1884            & 0.8793       &   -       &    -     &   -    \\ \hline
\end{tabular}
\end{table}
\begin{table}[ht]
\scriptsize
\centering
\caption{Simulation study: results based on 200 Monte Carlo samples with a different number of subjects. Maximum likelihood estimates, SEs, and absolute bias for the cases $b_{ij}\sim SN(0, 3, 4),\;  \bm{e}_{ij} \sim N_{pm} (\bm{0}, 0.7\bm{I}_{pm})$ and when both $\bm{e}_{ij}$, $b_{ij}$ are N. True parameter values are given in parentheses.}
\label{table simulated data1}
\begin{tabular}{lllllll} \hline
 & \multicolumn{3}{c}{If $\bm{e}_{ij}$ is N, $b_{ij}$ is SN} & \multicolumn{3}{c}{Both $\bm{e}_{ij}$ , $b_{ij}$ are N}  \\  \hline
Parameter                                        & Estimate           & SE               & $\left|\text{Bias}\right|$            & Estimate & SE     & $\left|\text{Bias}\right|$    \\ \hline
\multicolumn{7}{c}{$n_i$=30 for each sequence}  \\ \hline
$\beta_0$(3.3)                                & 3.3548             & 0.1023           & 0.1535          & 3.3597   & 0.1983 & 0.1584 \\
$\beta_{p_2}$(2.4)                                      & 2.4004             & 0.0515           & 0.0497          & 2.4004   & 0.0632 & 0.0497 \\
$\beta_{p_3}$(1.1)                                      & 1.1023             & 0.0610           & 0.0509          & 1.1023   & 0.0632 & 0.0509 \\
$\beta_{\tau_2}$(0.9)                                   & 0.8993             & 0.0617           & 0.0544          & 0.8993   & 0.0632 & 0.0544 \\
$\beta_{\tau_3}$(2.1)                                   & 2.0979             & 0.0545           & 0.0562          & 2.0979   & 0.0632 & 0.0562 \\
$\beta_{g_2}$(1.5)                                        & 1.5017             & 0.0693           & 0.0598          & 1.5017   & 0.0730 & 0.0598 \\
$\beta_{g_3}$(2.0)                                        & 2.0008             & 0.0663           & 0.0598          & 2.0008   & 0.0730 & 0.0598 \\
$\beta_{g_4}$(3.4)                                        & 3.3996             & 0.0513           & 0.0623          & 3.3996   & 0.0730 & 0.0623 \\
$\beta_1$(1.8)                                   & 1.7865             & 0.0580           & 0.1021          & 1.7846   & 0.1442 & 0.1132 \\
$\sigma_e^2$(0.7) & 0.7205             & 0.0110           & 0.0288          & 0.7205   & 0.8485 & 0.0288 \\
$\sigma_s^2$(3.0) & 2.9948             & 0.1923           & 0.5095          & 1.1948   & 1.0894 & 1.8052 \\
$\lambda_s$(4.0)                      & 4.2827             & 0.1771           & 1.2641          & -   & - & - \\ \hline
\multicolumn{7}{c}{$n_i$=50 for each sequence}                         \\ \hline
$\beta_0$(3.3)                                & 3.3617             & 0.0759           & 0.1124          & 3.3668   & 0.1495 & 0.1239 \\
$\beta_{p_2}$(2.4)                                      & 2.3965             & 0.0400           & 0.0380          & 2.3965   & 0.0491 & 0.0380 \\
$\beta_{p_3}$(1.1)                                      & 1.1003             & 0.0473           & 0.0378          & 1.1003   & 0.0491 & 0.0378 \\
$\beta_{\tau_2}$(0.9)                                   & 0.8993             & 0.0479           & 0.0433          & 0.8993   & 0.0491 & 0.0433 \\
$\beta_{\tau_3}$(2.1)                                   & 2.0973             & 0.0423           & 0.0414          & 2.0973   & 0.0491 & 0.0414 \\
$\beta_{g_2}$(1.5)                                        & 1.5051             & 0.0538           & 0.0463          & 1.5051   & 0.0567 & 0.0463 \\
$\beta_{g_3}$(2.0)                                        & 1.9985             & 0.0515           & 0.0482          & 1.9985   & 0.0567 & 0.0482 \\
$\beta_{g_4}$(3.4)                                        & 3.3974             & 0.0398           & 0.0418          & 3.3974   & 0.0567 & 0.0418 \\
$\beta_1$(1.8)                                   & 1.7868             & 0.0441           & 0.0750          & 1.7811   & 0.1105 & 0.0901 \\
$\sigma_e^2$(0.7) & 0.7230             & 0.0086           & 0.0190          & 0.7230   & 0.8502 & 0.0190 \\
$\sigma_s^2$(3.0) & 2.9621             & 0.1472           & 0.4180          & 1.1872   & 1.0867 & 1.8128 \\
$\lambda_s$(4.0)                      & 4.2618             & 0.1381           & 1.1209          & -   & - & - \\ \hline
\end{tabular}
\end{table}

 %Note in \cref{table simulated data} random effect is assumed to be normally distributed while in \cref{table simulated data1} random error is considered to be normally distributed.
 The results in \Cref{table simulated data} demonstrate that the estimates for bias and standard errors are lower when using the SN assumption, compared to assuming both the random error and random effect are normal. The decrease in bias and standard errors is more pronounced as the number of subjects per sequence increases to 50. Additionally, the proposed skew-normal model was deemed the most appropriate fit by 89\% of the AIC values when the actual random error was determined to be skew-normal, suggesting that the proposed algorithm is successful in handling skewed crossover data.
% From \Cref{table simulated data}, we notice the bias and standard errors of the estimates are lower in the case of SN assumption than in the case where both the random error and random effect are assumed to be normal. Standard errors and bias have further reduced when we increase the number of subject to 50 per sequence. When the actual random error is skew-normal, 89\% of AIC values select the proposed skew-normal model as the best fit.

 The results shown in \Cref{table simulated data1} indicate that the parameter estimates for period, treatment, and gene effects in the skew-normal model are comparable to those in the normal model, however, their standard errors differ. Specifically, the standard error for $\beta_{g_4}$ and $\beta_1$ in the skew-normal case is approximately half of that in the normal case. As previously reported by other researchers (\citet{lachos2010}), the estimate of the covariate $w_{ij}$'s parameter $\beta_1$ is impacted by both standard error and bias. Additionally, when the random effect is actually skew-normal, 83\% of AIC values suggest that the skew-normal fit is the best model, demonstrating that fitting standard model results in less accurate estimates for skewed crossover data.

\begin{figure}[ht]
\centering
  \includegraphics[width=9cm]{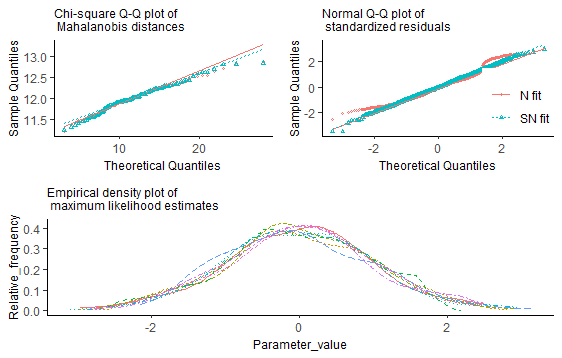}%{error skew residd.png}
  \caption{{Assesing model adequacy: a comparison of SN model  ($\bm{e}_{ij}$ is SN, and $b_{ij}$ is N) with the normal model for 30 subject sequences. The top panel compares the Mahalanobis distances to the theoretical $\chi^{2}_{12}$ distribution and the standardized residual vectors to a standard normal distribution. The `N fit' refers to a model fitted using normal random errors and random effects, while the `SN fit' refers to the proposed model. The bottom panel displays the estimated fixed effects through empirical density plots.}}
  \label{figerr}
 \end{figure}

\begin{figure}[H]
\centering
  \includegraphics[width=9cm]{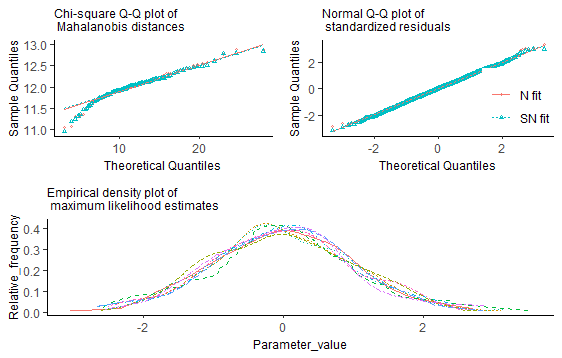}%{random skew resid.png}
  \caption{{Assesing model adequacy: a comparison of SN model  ($\bm{e}_{ij}$ is N and $b_{ij}$ is SN) with the normal model for 30 subject sequences. The top panel compares the Mahalanobis distances to the theoretical $\chi^{2}_{12}$ distribution and the standardized residual vectors to a standard normal distribution. The `N fit' refers to a model fitted using normal random errors and random effects, while the `SN fit' refers to the proposed model. The bottom panel displays the estimated fixed effects through empirical density plots.}}
  \label{figran}
 \end{figure}

The accuracy of the proposed SN model is evaluated by analyzing the chi-square Q-Q plot of the Mahalanobis distances ($\bm{d}_{ij}$) and the normal Q-Q plot of the standardized residual vectors ($\bm{r}_{ij}$). The top panels of Figures \ref{figerr} and \ref{figran} show that the quantiles of the proposed model are closely aligned with the diagonal reference line. Furthermore, the empirical density plot of the standardized maximum likelihood estimate is created, and the bottom panels of Figures \ref{figerr} and \ref{figran} indicate that they conform to the normal distribution curve. This leads us to the conclusion that the proposed model is appropriate for the data at hand.
%The overall evaluation of proposed SN model fitting is carried out by plotting the chi-square Q-Q plot of Mahalanobis distances $\bm{d}_{ij}$ and normal Q-Q plot standardized residual vectors $\bm{r}_{ij}$. We have also made the empirical density plot of estimated standardized maximum likelihood estimate and they resemble with normal curve. 
 %\textcolor{red}{(write a para on computation time and convergence aspects..., asymptotic normality of estimates and the residual plots for model validation, skewlmm package comparison in both cases and aic bic hq criteria)}
%%% REAL DATA ANALYSIS SECTION
\section{Gene Case Study Results}\label{gene sec}
 %In the exploratory analysis, we found that original and log transformed responses are asymmetric, period and treatment effects may be assumed to be the same for all the ten gene expression levels, and various interaction terms can be omitted. Also, fitting the gene data with the model having non-normal error distribution may provide a better solution. 
 In our preliminary investigation, we discovered that both the raw and log-transformed responses were skewed. We hypothesized that the period and treatment effects would be uniform across the ten gene expression levels and that specific interaction terms could be disregarded. Additionally, utilizing a model with a non-normal error distribution may result in a more favorable outcome for fitting the gene data.
 Thus, we fit the following statistical model with the underlying regression equation as given in \cref{eq:11}: 
\begin{align*}
y_{i j tk}&=\beta_{0}+\beta_{p_2} \text {Per}_{2}+\beta_{p_3} \text {Per}_{3}+\beta_{\tau_2} \mathrm{Trt}_{2}+\beta_{\tau_3} \mathrm{Trt}_{3}+\beta_{g_2} \text {Gene}_{2}\\&+\beta_{g_3} \text {Gene}_{3}+\beta_{g_4} \text {Gene}_{4}
+\beta_{g_5} \text {Gene}_{5}+\beta_{g_6} \text {Gene}_{6}+\beta_{g_7} \text {Gene}_{7}\\&+\beta_{g_8} \text {Gene}_{8}+\beta_{g_9} \text {Gene}_{9}+\beta_{g_{10}} \text {Gene}_{10}+s_{ij}
+e_{i j tk}, \numberthis \label{eq:11}
\end{align*}
where for the gene data $i,t=1,2,3;\, k=1(1)10$ and $j=1(1)n_i$, and $n_1=n_2=n_3=4$. Scenarios with different assumptions on $s_{i j}$ and $e_{i j tk}$ were used. The period, treatment, and gene effects were assumed to be fixed and represented by corresponding indicator variables.
The AIC and Bayesian information criterion (BIC) were used to compare the various cases. These cases are,
 \begin{itemize}
     \item[] Case 1: A model with normally distributed random error and random effects, \textit{i.e.,} when both $b_{ij}$ and $\bm{e}_{ij}$ are N.
     %are normally distributed %$s_{ij}\sim N(0, \sigma_s^2), \; e_{ijtk}\sim N(0,\sigma_e^2)$
     \item[] Case 2: A model with independent multivariate normal distribution for the random error and a univariate skew-normal distribution for random effects, \textit{i.e.,} when $b_{ij}$ is SN and $\bm{e}_{ij}$ is N.
     \item[] Case 3: A model with independent multivariate skew-normal distribution for random error and univariate normal distribution for random effects, \textit{i.e.,}  when $b_{ij}$ is N and $\bm{e}_{ij}$ is SN.
 \end{itemize}
Table \ref{table real data} reports the maximum likelihood estimates and estimated asymptotic standard errors of the parameters $(\bm{\beta},\sigma_{e}^2,\sigma_{s}^2,\lambda)$ for the three cases. As an initial value of mean and variance parameters, estimates obtained by fitting a normal linear mixed effect model were used in the EM algorithm. %However, an estimate of skewness was assumed to be in the range of $0.001$. 
From \Cref{table real data}, we observe that parameter estimates and standard errors of Case 2 are close to the estimates obtained under the normality assumption (Case 1), implying that the asymmetry is not detected in random effects. However, Case 3 suggests the best fit based on AIC and BIC values, supporting the argument of departure from the normality of the residuals. %The estimates and SE of period effects and individual-level covariates also differ in model 3, supporting the findings of simulations.

\begin{figure}
\centering
  \includegraphics[width=9cm]{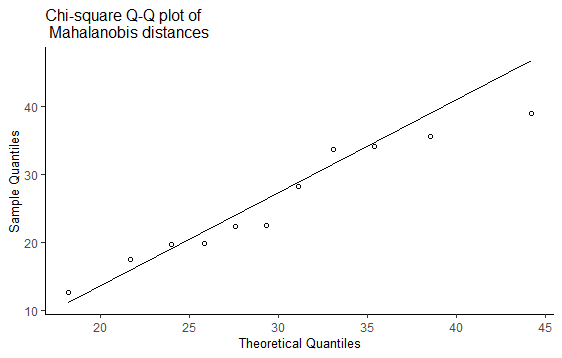}
  \caption{{Model fit comparison for Case 3: Mahalanobis Distances \textit{vs.} Theoretical $\chi^{2}_{30}$ Distribution.}}
  \label{figd}
 \end{figure}

The adequacy of the selected model in Case 3 is assessed by examining the Mahalanobis distances $\bm{d_{ij}}$. %and the residual vectors $\bm{r}_{ij}$. 
A Kolmogorov–Smirnov (KS) test (\citet{kstest1951}) is most commonly used to test the goodness of fit of data to a theoretical distribution. %The Kolmogorov–Smirnov (KS) test is a nonparametric goodness-of-fit test and is used to determine whether two distributions differ or whether an underlying probability distribution differs from a hypothesized distribution. 
KS test of Mahalanobis distances $\bm{d_{ij}}$ gives a p-value of 0.06, which indicates the calculated distances are generated from the chi-square distribution at the 5\% significance level. %The KS test of the residual vector $\bm{r}_{ij}$ results in a p-value of 0.02, which indicates residuals are derived from normal distributions, assuming a 1\% significance level. 
Furthermore, we have constructed chi-square Q-Q plots for the Mahalanobis distances $\bm{d_{ij}}$ by which we can compare the observed and expected values of the Mahalanobis distances.
%We have also constructed chi-square Q-Q plots of Mahalanobis distances $\bm{d_{ij}}$ and normal Q-Q plots of residual vectors $\bm{r}_{ij}$.
%In addition, we have constructed chi-square Q-Q plots of Mahalanobis distances $\bm{d_{ij}}$ and normal Q-Q plots of residual vectors $r_{ij}$ as measures of model adequacy.
%The Q-Q plots allow us to compare the observed and expected values of the Mahalanobis distances and residual vectors. 
\Cref{figd} shows that the observed values are close to the expected values, indicating that the model defined in Case 3 is appropriately fitted. %Ks test also gives p-value 0.06 for di and 0.02 for rij
%Furthermore, as the model adequacy, we have made the chi-square Q-Q plot of Mahalanobis distances  and normal Q-Q plot of residual vector $r_{ij}$.  indicates the 
 %in the left plot and standardized residuals \textit{vs.} standard normal distribution in the right plot.
\begin{table}[ht]
\centering
\caption{Analysis of gene data: Maximum likelihood estimates, SEs, based on
fitting cases 1 to 3. Comparing cases 1, 2, and 3 using AIC and BIC values.
}
\scriptsize
\label{table real data}
\begin{tabular}{cllllll}
\hline
 & \multicolumn{2}{c}{Case 1 (both $b_{ij}$ , $\bm{e}_{ij}$ is N)}   & \multicolumn{2}{c}{ Case 2 ($b_{ij}$ is SN, $\bm{e}_{ij}$ is N)} & \multicolumn{2}{c}{Case 3 ($b_{ij}$ is N, $\bm{e}_{ij}$ is SN)}              \\ 
Parameter & Estimate & SE     & Estimate& SE                 & Estimate& SE  \\ \hline
$\beta_0$         & 3.4896     & 0.022  & 3.4896    & 0.0187 & 3.4486    & 0.0069 \\
$\beta_{p_2}$   & 0.0007     & 0.0138 & 0.0007    & 0.0138 & 0.0117    & 0.0128               \\
$\beta_{p_3}$     & 0.0118     & 0.0138 & 0.0118    & 0.0138 & 0.0228    & 0.0127                \\
$\beta_{\tau_2}$  & -0.0055    & 0.0138 & -0.0055   & 0.0138 & -0.0088   & 0.0117            \\
$\beta_{\tau_3}$  & 0.0081     & 0.0138 & 0.0081    & 0.0138 & 0.0016    & 0.0118\\
$\beta_{g_2}$     & -0.9284    & 0.0252 & -0.9284   & 0.0239 & -0.8915   & 0.0188            \\
$\beta_{g_3}$     & -1.5865    & 0.0252 & -1.5865   & 0.021  & -1.5496   & 0.0173               \\
$\beta_{g_4}$     & -0.3275    & 0.0252 & -0.3275   & 0.0251 & -0.2906   & 0.0182               \\
$\beta_{g_5}$     & -0.0381    & 0.0252 & -0.0381   & 0.0252 & -0.0012   & 0.0171                \\
$\beta_{g_6}$    & -1.2646    & 0.0252 & -1.2646   & 0.0226 & -1.2278   & 0.0183                \\
$\beta_{g_7}$     & -1.4755    & 0.0252 & -1.4755   & 0.0216 & -1.4386   & 0.0177              \\
$\beta_{g_8}$    & -2.0328    & 0.0252 & -2.0328   & 0.0178 & -1.9959   & 0.0146               \\
$\beta_{g_9}$     & -0.8211    & 0.0252 & -0.8211   & 0.0242 & -0.7842   & 0.0188                      \\
$\beta_{g_{10}}$   & -0.6482    & 0.0252 & -0.6482   & 0.0246 & -0.6113   & 0.0187                      \\
   $\sigma_e^2$           & 0.0115     & 0.0219 & 0.0115    & 0.0001 & 0.0117    & 0.0001                     \\
   $\sigma_s^2$               & 0.0005     & 0.107  & 0.0005    & 0.0004      & 0.0003    & 0.0002                  \\
    $\lambda$              & -        & -     & 0.001     & 0.3856 & 3.8849    & 0.0496           \\ \hline
AIC                                          & \multicolumn{2}{c}{-545.50} & \multicolumn{2}{c}{-528.08}               & \multicolumn{2}{c}{-553.64}       \\
BIC                             & \multicolumn{2}{c}{-483.32} & \multicolumn{2}{c}{-465.90}               & \multicolumn{2}{c}{-491.46}                           \\
\hline
\end{tabular}
\end{table}
%\section{Concluding remarks}
\section{Computational Details}\label{comp}
For simulation and real data analyses, R programming (\citet{rr}) has been used. We used R version 4.1.0 under Windows 10 (64-bit), with an Intel core I5 processor and 4GB of RAM. 

The simulation specifications outlined in section \Cref{gene sec} were utilized to create 100 simulated data sets, each containing 30 subjects per sequence. It took approximately 1.8 hours to run the simulation when $b_{ij}$ was designated as SN and $\bm{e}_{ij}$ was N. The EM algorithm converged in an average of 45-50 steps in these data sets. In the scenario where $b_{ij}$ was designated as N and $\bm{e}_{ij}$ was SN, the simulation took approximately 45 minutes to run and the EM algorithm converged in an average of 15-20 steps. The normal linear mixed model was then fit to the data using the `lme' function from the \textit{nlme} library in R. % The within-subject dependence structure was considered when estimating the model parameters of the skew-normal linear mixed model with the `smsn.lmm' function of the \textit{skewlmm} library. % Existing methods were used to analyze the skewed data using R functions. 

The gene data exploratory analysis (as shown in \Cref{case study}) and model adequacy plots (as shown in \Cref{sim study}) were depicted through Q-Q plots, and empirical density plots using the \textit{ggplot2}, \textit{ggpubr}, and \textit{tidyverse} libraries. In gene data analysis, the EM algorithm converged in 2 steps for Case 2, where $b_{ij}$ is SN and $\bm{e}_{ij}$ is N, whereas 14 steps were required for convergence in Case 3, where $b_{ij}$ is N and $\bm{e}_{ij}$ is SN. Using the R platform with similar settings, a gene data model can be fitted in approximately 4-5 minutes. The R programs used are available at the provided hyperlink, \href{https://github.com/savitapareek/Likelihood-based-Inference-for-Skewed-Responses-in-a-Crossover-Trial-Setup.git}{rprograms}.
%\section{Concluding remarks}
\section{Concluding Remarks}\label{conc}
We have proposed an EM algorithm-based estimation for multiple and skewed crossover data by moderating the assumptions on random effect and model error densities. Closed-form expression are obtained for fixed effects, and variance and skewness components are obtained using the first-order NR equation. A small simulation study is carried out to highlight the potential gain in the efficiency of some parameters when the normality assumption does not hold, with some extra computing cost. 
%A number of studies have employed skew-normal/independent linear mixed models to analyze longitudinal data with skewed responses (\citet{lachos2010}, \citet{schu2020}). In a skew linear mixed model, the random effects are assumed to be a scale mixture of skew-normal distributions, while the random errors are assumed to be a scale mixture of normal distributions. In our article, crossover data is analyzed using a linear mixed model, and asymmetry in the data can be accounted for by the skew normality of random effects or random errors in the model. The proposed parameter estimation technique has been compared with existing tools through simulation studies. 

The approach used in this paper can be used in treating other multivariate models, such as having different treatment, period, and gene effects for each response variate. 
We believe this idea is applicable by assuming both the random effect and random error to be skew normally distributed and having the general variance-covariance structure. It is also possible to consider dependence structures in random errors, such as damped exponential correlations and serially autoregressive correlations of order $p$. As a consequence, the estimation problem becomes more complex. One can also apply the proposed technique to study other classes of asymmetric distributions such as skew-t, skew-slash, and skew-contaminated normal. %We have implemented our approach using R programming; code is available from the first author upon request. 

\vspace*{0.6pc}
\noindent {\bf{Acknowledgement}}
We thank Dr. Atanu Bhattacharjee, Tata Memorial Center, Mumbai, India, for his assistance in obtaining the gene data set.
\vspace*{0.6pc}
\bibliographystyle{plainnat}
\bibliography{libr}

\section{Appendix}\label{sec appen}
\subsection{Skew-Normal Distribution}\label{UN SN}
The following is a brief overview of the skew-normal distribution and the terminology that we have used in our analysis.
%We define the univariate and multivariate skew normal distribution and the parameters that will be used in the following sections.
\begin{itemize}
    \item[(i)] Univariate skew-normal variate (\citet{azzalini1985}): If a random variable $W$ has the density function
    \begin{equation*}
    f(w;\lambda)=2\phi(w)\Phi(\lambda w),\;\; -\infty <w< \infty.
    \end{equation*}
    where $\phi$, $\Phi$ are the standard normal density and distribution function, respectively, then we say $W$ is a skew-normal variate with parameter $\lambda$, or, $W \sim SN(0,1,\lambda)$. The random variable $W$ has additive representation in terms of normal and half normal distribution, \textit{i.e.,} if $U_0, U_1$ are independent standard normals, $\delta \in (-1, 1)$ and `$\overset{d}{=}$' meaning is `distributed as' then $$W\overset{d}{=}\sqrt{1-\delta^2}U_0 +\delta |U_1|, \; \; \delta=\frac{\lambda}{\sqrt{(1+\lambda^2)}}.$$ 
    
   The parameters $(0, 1, \lambda)$ are not the true mean, variance, and skewness of the random variate $W$. The moments of skew-normal can be found from its additive representation as  \begin{equation}
       \mu_w= \delta  \sqrt{2/ \pi} , \;\; \sigma^2_w= 1- \delta^2 (2/ \pi), \;\; \gamma_1(W)=\frac{4-\pi}{2}\frac{\mu_w^3}{\sigma_w^3  }.\label{momu}
   \end{equation}
    
For applied work, we need to introduce location and scale parameters. If $W_1$ is a continuous random variable with location and scale parameters $\xi, \omega$, respectively then the variable $W_1=\xi+\omega W$ will be skew-normal with parameters $\xi, \omega^2, \lambda$, with density function $$f_{W_1}(w_1)=2\phi(w_1|\xi,\omega^2)\Phi\left(\lambda \frac{w_1-\xi}{\omega}\right), \; -\infty <w_1< \infty.$$
We use the notation $W_1 \sim SN(\xi, \omega^2, \lambda)$. When $\lambda =0$, it reduces to the normal distribution. Skewed-normal distributions, which are parametric families governed by moments up to third order, may need a larger sample size than the traditional rule of thumb of n=30 (\citet[Chapter~3]{azzalini2014}). 

The parameters $(\xi, \omega^2, \lambda)$ are called direct parameters (DP) as they appear in the density function and are used to regulate the mean, variance, and skewness parameters. However, the expected and observed Fisher information matrix is singular when $\lambda$ is close to 0 (\citet{azzalini2014}). \citet{azzalini1985} proposed that MLE inference be based on centered parameters (CP) to overcome the singularity problem. CP parameters $(\mu, \sigma^2, \gamma)$ represent the actual mean, variance, and skewness parameters that are derived from an additive representation (\cref{momu}). %In the regression setting, only the intercept has to be adjusted. 
CP parameters $(\mu, \sigma^2, \gamma)$ are more familiar than DP parameters $(\xi, \omega^2, \lambda)$, which makes their interpretation easy. Because CP parameters are interpretable and close to asymptotic normal distributions, they are preferred over DP parameters. As a result, CP parameters are suitable for constructing confidence intervals and other inference methods (\citet{azzalini1999}).
%The parameters $(\xi, \omega^2, \lambda)$ are called the direct parameters (DP) as they appear in the density function and regulate the mean, variance, and skewness parameters. The actual mean, variance, and skewness parameter $(\mu, \sigma^2, \gamma)$, also known as CP parameters, are obtained from the additive representation (\cref{momu}).
More properties of this distribution can be found in \citet{genten2005}.
  \item[(ii)] Multivariate skew-normal variate (\citet{lachos2005}): An $n$ dimensional random vector $\bm{W}_1$ follows a skew-normal distribution with location vector $\bm{\mu} \in \mathbb{R}^{n}$, a dispersion matrix $\bm{\Sigma}$ (a $n \times n$ positive definite matrix) and skewness vector $\bm{\lambda} \in \mathbb{R}^{n}$, if its pdf is given by 
  \begin{equation*}
      f_{\bm{W}_1}(\bm{w_1})=2\phi_n(\bm{w}_1|\bm{\mu},\bm{\Sigma})\Phi_1(\bm{\lambda}^{T}\bm{\Sigma}^{-1/2}(\bm{w}_1-\bm{\mu})),\;\; \bm{w}_1 \in \mathbb{R}^{n}.
  \end{equation*} 
  We denote it by $\bm{W}_1\sim SN_n(\bm{\mu},\bm{\Sigma},\bm{\lambda})$, further $\bm{W}_1$ can be written as $\bm{W}_1{=} \bm{\mu}+\bm{\Sigma}^{1/2}\bm{W}$, where, $\bm{W}$ is a standardized skew-normal vector. This definition is derived from the fundamental skew-normal distribution introduced by \citet{azzalini1996}, \citet{azzalini1999}. The standardised skew-normal, $\bm{W} \sim SN (\bm{0}, \bm{I},\bm{\lambda})$ can also be represented as 
  $$\bm{W} \overset{d}{=} \bm{\delta}|U_0|+(\bm{I}_n-\bm{\delta}\bm{\delta}^T)^{1/2}\bm{U}_1, \;\; \bm{\delta}=\frac{\bm{\lambda}}{\sqrt{1+\bm{\lambda}^T \bm{\lambda}}},$$
 where, $U_0 \sim N(0,1)$ independent of $\bm{U}_1 \sim N_n(\bm{0},\bm{I}_n)$. One important aspect of the multivariate skew normal is that the joint independence of a random vector holds only if, at most, one of them is marginally skew-normal (\citet{azzalini2014}). In \citet{azzalini2008}, a centered parameterization (CP) is proposed for multivariate SN distributions. 
\end{itemize}
\subsection{Some Results for Maximising the Likelihood}\label{appres}
We will use the following results from \citet{lachos2005} to implement the two steps of EM for maximizing the likelihood.
\begin{lemma}\label{res 1}
Let $\textbf{W} \sim {SN}_{n}({\bm{\lambda}})$.
Then $$\textbf{W} \overset{d}{=}\bm{\delta}|X_{0}|+(\textbf{I}_{n}-\bm{\delta}\bm{\delta}^{T})^{1/2}\textbf{X}_{1}, \;\; \text{where} \; \bm{\delta}=\frac{\bm{\lambda}}{\sqrt{1+\bm{\lambda^{T}}\bm{\lambda}}},$$
$X_{0}\sim N_{1}(0,1)$ independent of $\textbf{X}_{1} \sim N_{n}(\textbf{0},\textbf{I}_{n})$ and `$\overset{d}{=}$' meaning `distributed as'.
\end{lemma}
\begin{lemma}\label{res 2}
Suppose that $\bm{Y}|T=t \sim N_{n}(\bm{\mu}+\bm{d}t,\bm{\Psi})$ and $T \sim HN_{1}(0,1)$ (the standardized half normal distribution). Let $\bm{\Sigma}=\bm{\Psi}+\bm{d}\bm{d}^{T}$. Then the joint distribution of $(\bm{Y}^{t},T)^{T}$ can be written as
$$f_{\bm{Y},T}(\bm{y},t|\bm{\theta,\lambda})= 2\phi_{n}(\bm{y|\mu,\Sigma})\phi_{1}(t|\eta,\zeta^{2})I\{t>0\}.$$  The marginal distribution of $\bm{Y}$ after integrating out $t$ is given by $$f_{\bm{Y}}(\bm{y}|\bm{\theta,\lambda})= 2\phi_{n}(\bm{y|\mu,\Sigma})\Phi_{1}(\eta|\zeta),$$ 
 $$\text{where,}\;\eta= \frac{\bm{d}^{T}\bm{\Psi}^{-1}(\bm{y}-\bm{\mu})}{1+\bm{d}^{T}\bm{\Psi}^{-1}\bm{d}}\;\; \text{and} \; \zeta^{2}=\frac{1}{1+\bm{d}^{T}\bm{\Psi}^{-1}\bm{d}}.$$
\end{lemma} 
\begin{lemma}\label{res 3}
 Under the condition in \Cref{res 2}, $$E[T^{k}|\bm{y}]=E[X^{k}|X>0],$$ where $X \sim N_{1}(\eta, \zeta ^2)$ with $\eta$ and $\zeta ^2$ given above. Particularly,$$E[T|\bm{y}]=\eta+\frac{\phi(\frac{\eta}{\zeta})}{\Phi_{1}(\frac{\eta}{\zeta})}\zeta,$$
$$\text{and}\; E[T^{2}|\bm{y}]=\eta^{2}+\zeta^{2}+\frac{\phi(\frac{\eta}{\zeta})}{\Phi_{1}(\frac{\eta}{\zeta})}\eta\zeta.$$
\end{lemma}
\subsection{Elements of Hessian Matrix for SN Errors}\label{rerr}
Following the notation defined in \Cref{error SN}, we proceed as follows to find the second-order derivative of the Q-function.
For $a=1(1)3, b=a(1)3 $, we have,
\begin{align*}
\frac{\partial^2 Q}{\partial {\xi}_{[a]}\partial{\xi}_{[b]}}&=\sum_{i=1}^{s}\sum_{j=1}^{n_{i}}-\frac{1}{2}[\text{tr}(\bm{V}_{ij{\xi}_{[b]}}^{*}\bm{V}_{ij{\xi}_{[a]}}+\bm{V}_{ij}^{-1}\bm{V}_{ij{\xi}{\xi}_{[a,b]}})\\
&+\bm{d}_{ij}^T {\bm{V}^{*}_{ij{\xi}\xi_{[a,b]}}}\bm{d}_{ij} T_{ij}^{02}+2\bm{d}_{ij}^T{\bm{V}^{*}_{ij{\xi}_{[a]}}}{\bm{d}_{ij{\xi}_{[b]}}} T_{ij}^{02}
\\&+2(\bm{d}_{ij}^T\bm{V}_{ij}^{-1}{\bm{d}_{ij{\xi}{\xi}_{[a,b]}}}+{\bm{d}_{ij{\xi}_{[b]}}}\bm{V}_{ij}^{-1}{\bm{d}_{ij{\xi}_{[a]}}}+\bm{d}_{ij}^T\bm{V}_{ij{\xi}_{[b]}}^{*}{\bm{d}_{ij{\xi}_{[a]}}}) T_{ij}^{02}\\
&+(\bm{y}_{ij}-\bm{X}_{ij}\bm{\beta})^{T}{\bm{V}^{*}_{ij{\xi}\xi_{[a,b]}}}(\bm{y}_{ij}-\bm{X}_{ij}\bm{\beta}-2\bm{d}_{ij}T_{ij}^{01})\\
&-2(\bm{y}_{ij}-\bm{X}_{ij}\bm{\beta})^{T}{\bm{V}^{*}_{ij{\xi}_{[a]}}}{\bm{d}_{ij{\xi}_{[b]}}}T_{ij}^{01}
-2(\bm{y}_{ij}-\bm{X}_{ij}\bm{\beta})^{T}{\bm{V}^{*}_{ij{\xi}_{[b]}}}{\bm{d}_{ij{\xi}_{[a]}}}T_{ij}^{01}\\
&-2(\bm{y}_{ij}-\bm{X}_{ij}\bm{\beta})^{T}{\bm{V}^{-1}_{ij}}{\bm{d}_{ij{\xi}{\xi}_{[a,b]}}}T_{ij}^{01}],
\end{align*}
where, \begin{align*}
\bm{V}_{ij{\xi}{\xi}}&=\frac{\partial\bm{V}_{ij\bm{\xi}}}{\partial\bm{{\xi}}}=(\bm{V}_{ij{\xi}{\xi}_{[a,b]}})_{3\times 3}
=\begin{bmatrix}
\frac{\partial^2 \bm{V}_{ij}}{\partial\sigma_e^2\partial \sigma_e^2} & \frac{\partial^2 \bm{V}_{ij}}{\partial\sigma_e^2 \partial\sigma_s^2}  &\frac{\partial^2 \bm{V}_{ij}}{\partial\sigma_e^2 \partial\lambda_e}   \\
 & \frac{\partial^2 \bm{V}_{ij}}{\partial\sigma_s^2 \partial\sigma_s^2} & \frac{\partial^2 \bm{V}_{ij}}{\partial\sigma_s^2 \partial\lambda_e}\\
 & & \frac{\partial^2 \bm{V}_{ij}}{\partial\lambda_e\partial \lambda_e} 
\end{bmatrix},\\
\frac{\partial^2 \bm{V}_{ij}}{\partial\sigma_e^2 \partial\lambda_e}&= \bm{R}_{l},\;\;
\frac{\partial^2 \bm{V}_{ij}}{\partial\lambda_e\partial \lambda_e}=\sigma_e^2 \bm{R}_{ll}, \\
\bm{R}_{ll}&=\frac{\partial \bm{R}_l}{\partial \lambda}=-2(\bm{\delta}_e\bm{\delta}_{ell}^{T}+\bm{\delta}_{el}\bm{\delta}_{el}^{T}),\\
{\bm{V}^{*}_{ij{\xi}\xi_{[a,b]}}}&=\frac{\partial^2 \bm{V}_{ij}^{-1}}{\partial{\xi}_{[a]}\partial{\xi}_{[b]}}=-\bm{V}_{ij}^{-1}\bm{V}_{ij{{\xi}{\xi}}_{[a,b]}}\bm{V}_{ij}^{-1}-2\bm{V}_{ij{\xi}_{[b]}}^{*}\bm{V}_{ij{{\xi}_{[a]}}}\bm{V}_{ij}^{-1}.
\end{align*}
Remaining entries of the matrix $\bm{V}_{ij{\xi}{\xi}}$ are zero.
\begin{align*}
\bm{d}_{ij{\xi}{\xi}}&=\frac{\partial\bm{d}_{ij\bm{\xi}}}{\partial\bm{{\xi}}}=(\bm{d}_{ij{\xi}{\xi}_{[a,b]}})_{3\times 3}
=\begin{bmatrix}
 \frac{\partial^2 \bm{d}_{ij}}{\partial\sigma_e^2 \partial\sigma_e^2}&0&\frac{\partial^2 \bm{d}_{ij}}{\partial\sigma_e^2 \partial\lambda_e}\\
&0&0\\
& & \frac{\partial^2 \bm{d}_{ij}}{\partial\lambda_e\partial \lambda_e}
\end{bmatrix},\\
\frac{\partial^2 \bm{d}_{ij}}{\partial\sigma_e^2 \partial\sigma_e^2}&=\frac{-\bm{\delta}_{e}}{4 {\sigma_e}^{3}},\;\;
\frac{\partial^2 \bm{d}_{ij}}{\partial\sigma_e^2 \partial\lambda_e}=\frac{\bm{\delta}_{el}}{2 \sqrt{\sigma_e^2}},\\
\frac{\partial^2 \bm{d}_{ij}}{\partial\lambda_e\partial \lambda_e}&=\sigma_e \bm{\delta}_{ell},\;\;
\bm{\delta}_{ell}=\frac{\partial \bm{\delta}_{el}}{\partial \lambda_e}=\frac{-3\lambda_e (1,0,\ldots,0)}{(1+\lambda_e^2)^{5/2}}.
\end{align*}
\subsection{Elements of Hessian Matrix for SN Random Effect}\label{rskew}
Based on the notation defined in \Cref{random SN}, we proceed as follows to find the second order derivative of the Q-function.
For $a=1(1)3, b=a(1)3 $, we have,
\begin{align*}
\frac{\partial^2 Q}{\partial {\xi}_{[a]}\partial{\xi}_{[b]}}&=\sum_{i=1}^{s}\sum_{j=1}^{n_{i}}-\frac{1}{2}[\text{tr}(\bm{V}_{ij{\xi}_{[b]}}^{*}\bm{V}_{ij{\xi}_{[a]}}+\bm{V}_{ij}^{-1}\bm{V}_{ij{\xi}{\xi}_{[a,b]}})\\
&+\bm{d}_{ij}^T {\bm{V}^{*}_{ij{\xi}\xi_{[a,b]}}}\bm{d}_{ij} T_{ij}^{02}+2\bm{d}_{ij}^T{\bm{V}^{*}_{ij{\xi}_{[a]}}}{\bm{d}_{ij{\xi}_{[b]}}} T_{ij}^{02}
\\&+2(\bm{d}_{ij}^T\bm{V}_{ij}^{-1}{\bm{d}_{ij{\xi}{\xi}_{[a,b]}}}+{\bm{d}_{ij{\xi}_{[b]}}}\bm{V}_{ij}^{-1}{\bm{d}_{ij{\xi}_{[a]}}}+\bm{d}_{ij}^T\bm{V}_{ij{\xi}_{{[b]}}}^{*}{\bm{d}_{ij{\xi}_{[a]}}}) T_{ij}^{02}\\
&+(\bm{y}_{ij}-\bm{X}_{ij}\bm{\beta})^{T}{\bm{V}^{*}_{ij{\xi}\xi_{[a,b]}}}(\bm{y}_{ij}-\bm{X}_{ij}\bm{\beta}-2\bm{d}_{ij}T_{ij}^{01})\\
&-2(\bm{y}_{ij}-\bm{X}_{ij}\bm{\beta})^{T}{\bm{V}^{*}_{ij{\xi}_{[a]}}}{\bm{d}_{ij{\xi}_{[b]}}}T_{ij}^{01}
-2(\bm{y}_{ij}-\bm{X}_{ij}\bm{\beta})^{T}{\bm{V}^{*}_{ij{\xi}_{[b]}}}{\bm{d}_{ij{\xi}_{[a]}}}T_{ij}^{01}\\
&-2(\bm{y}_{ij}-\bm{X}_{ij}\bm{\beta})^{T}{\bm{V}^{-1}_{ij}}{\bm{d}_{ij{\xi}{\xi}_{[a,b]}}}T_{ij}^{01}],
\end{align*}
where, \begin{align*}
\bm{V}_{ij{\xi}{\xi}}&=\frac{\partial\bm{V}_{ij\bm{\xi}}}{\partial\bm{{\xi}}}=(\bm{V}_{ij{\xi}{\xi}_{[a,b]}})_{3\times 3}
=\begin{bmatrix}
\frac{\partial^2 \bm{V}_{ij}}{\partial\sigma_e^2\partial \sigma_e^2} & \frac{\partial^2 \bm{V}_{ij}}{\partial\sigma_e^2 \partial\sigma_s^2}  &\frac{\partial^2 \bm{V}_{ij}}{\partial\sigma_e^2 \partial\lambda_s}   \\
 & \frac{\partial^2 \bm{V}_{ij}}{\partial\sigma_s^2 \partial\sigma_s^2} & \frac{\partial^2 \bm{V}_{ij}}{\partial\sigma_s^2 \partial\lambda_s}\\
 & & \frac{\partial^2 \bm{V}_{ij}}{\partial\lambda_s\partial \lambda_s} 
\end{bmatrix},\\
\frac{\partial^2 \bm{V}_{ij}}{\partial\sigma_s^2 \partial\lambda_s}&=R_l \bm{Z}_{ij}\bm{Z}_{ij}^{T},\;\;
\frac{\partial^2 \bm{V}_{ij}}{\partial\lambda_s\partial \lambda_s}=\sigma_s^2 R_{ll} \bm{Z}_{ij}\bm{Z}_{ij}^{T},\\
R_{ll}&=\frac{\partial R_l}{\partial \lambda_s}=-2(\delta_b\delta_{bll}+\delta_{bl}^2),\\
{\bm{V}^{*}_{ij{\xi}\xi_{[a,b]}}}&=\frac{\partial^2 \bm{V}_{ij}^{-1}}{\partial{\xi}_{[a]}\partial{\xi}_{[b]}}=-\bm{V}_{ij}^{-1}\bm{V}_{ij{{\xi}{\xi}}_{[a,b]}}\bm{V}_{ij}^{-1}-2\bm{V}_{ij{\xi}_{[b]}}^{*}\bm{V}_{ij{{\xi}_{[a]}}}\bm{V}_{ij}^{-1}.
\end{align*}
Remaining entries of the matrix $\bm{V}_{ij{\xi}{\xi}}$ are zero.
\begin{align*}
\bm{d}_{ij{\xi}{\xi}}&=\frac{\partial\bm{d}_{ij\bm{\xi}}}{\partial\bm{{\xi}}}=(\bm{d}_{ij{\xi}{\xi}_{[a,b]}})_{3\times 3}
=\begin{bmatrix}
0&0&0\\
& \frac{\partial^2 \bm{d}_{ij}}{\partial\sigma_s^2 \partial\sigma_s^2}&\frac{\partial^2 \bm{d}_{ij}}{\partial\sigma_s^2 \partial\lambda_s}\\
& & \frac{\partial^2 \bm{d}_{ij}}{\partial\lambda_s\partial \lambda_s}
\end{bmatrix},\\
\frac{\partial^2 \bm{d}_{ij}}{\partial\sigma_s^2 \partial\sigma_s^2}&=\frac{-\bm{Z}_{ij}\delta_{b}}{4 {\sigma_s}^{3}},\;\;
\frac{\partial^2 \bm{d}_{ij}}{\partial\sigma_s^2 \partial\lambda_s}=\frac{\bm{Z}_{ij}\delta_{bl}}{2 \sqrt{\sigma_s^2}},\\
\frac{\partial^2 \bm{d}_{ij}}{\partial\lambda_s\partial \lambda_s}&=\bm{Z}_{ij}\sigma_s \delta_{bll},\;\;
\delta_{bll}=\frac{\partial \delta_{bl}}{\partial \lambda_s}=\frac{-3\lambda_s}{(1+\lambda_s^2)^{5/2}}.
\end{align*}




\end{document}